\documentclass[12pt,epsfig]{article}  
\usepackage{amsmath}
\usepackage{amssymb}
\usepackage{graphicx}
\usepackage{subfig}
\usepackage{commath}
\usepackage[left=2.0cm,right=2.0cm, top=2.5cm,bottom=2.5cm]{geometry}
\usepackage{cite}

\begin{document}

\newcommand{\sheptitle}
{Mu-Tau Reflection Symmetry in the Standard Parametrization and Contributions
from Charged Lepton Sector}
\newcommand{\shepauthor}
{Chandan Duarah \footnote{ E-mail: chandan.duarah@gmail.com}}
\newcommand{\shepaddress}
   { Department of Physics, Dibrugarh University,
               Dibrugarh - 786004, India }
\newcommand{\shepabstract}
{The $\mu-\tau$ reflection symmetry of the lepton mixing matrix accommodates maximal
atmospheric mixing ($\theta_{23}=\pi/4$) as well as maximal Dirac CP phase 
($\delta=\pm \pi/2$) for the Dirac case. In the standard parametrization of the PMNS 
matrix the reflection symmetric nature is not directly visible while substituting 
the maximal values of the mixing parameters. This issue has been addressed in this paper. 
It is found that the reflection symmetry in the 'standard' PMNS matrix can be 
restored by allowing maximal values of the Majorana CP phases ($\alpha$, $\beta$)
as well, along with maximal $\delta$. To accommodate non-maximal values of 
$\theta_{23}$ and $\delta$ we consider charged lepton contributions to the neutrino 
mixing and implement the proposed scheme of reflection symmetry in the neutrino
mixing matrix. The charged lepton correction scheme succeeds in the prediction of   
lepton mixing parameters consistent with the global $3\nu$ 
oscillation data.\\

Key-words: Lepton mixing, $\mu-\tau$ reflection symmetry, charged lepton contributions.\\
\indent PACS number: 14.60 Pq}
\begin{titlepage}
\begin{flushright}
\end{flushright}
\begin{center}
{\large{\bf\sheptitle}}
\bigskip\\
\shepauthor
\\
\mbox{}\\
{\it\shepaddress}\\
\vspace{.5in}
{\bf Abstract}
\bigskip
\end{center}
\setcounter{page}{0}
\shepabstract
\end{titlepage}


\section{Introduction}
\indent The measurement of the non-zero reactor angle $\theta_{13}$ \cite{dbay,reno,dchoo} 
elevates neutrino 
physics research one step ahead. It also initiates the exploration of leptonic CP violation in 
oscillation experiments. The Dirac CP violating phase $\delta$ is likely to be determined soon 
with good accuracy whereas the problems of octant degeneracy and mass ordering still require
their solutions. Recent data from T2K \cite{t2k}, NO$\nu$A \cite{nova} and IceCube \cite{icube} 
experiments indicates a preference for the atmospheric angle $\theta_{23}$ to lie in the second 
octant which is also reflected in the global analysis of neutrino oscillation data made in 
Refs.\cite{global,global1}. The global analysis also indicates 
that the value of $\delta$ is 
close to $-\pi/2$. \\
\indent The approximate mixing pattern revealed by the oscillation data stems the main motivation 
towards $\mu-\tau$ flavour symmetry to understand the theory of lepton mixing. The near maximal 
value of the atmospheric mixing angle $\theta_{23}$ predicted by oscillation data is the key 
point behind $\mu-\tau$ flavour symmetry. The flavour symmetry, mostly exercised in lepton 
flavour models, is the so called $\mu-\tau$ permutation 
symmetry. It accommodates the well known predictions- $\theta_{23}=\pi/4$ and $\theta_{13}=0$, 
that dominates the field of neutrino physics research over a long period of time. 
The permutation symmetry embedded with a CP conjugation of the lepton sector is referred 
to as the $\mu-\tau$ reflection symmetry. This concept of $\mu-\tau$ reflection symmetry
was first put forwarded by Harrison and Scott \cite{hs}. Subsequently the mass matrix bearing 
the reflection symmetry property is realized in a $A_4$ based model by Babu, Ma and Valle 
\cite{bmv} and a general treatment of the reflection symmetry is rendered in Ref.\cite{grimus}. 
A review of $\mu-\tau$ flavour symmetry is also available in Ref.\cite{xing}. \\
\indent The prediction $\theta_{23}=\pi/4$ is a common feature of $\mu-\tau$ symmetry and
of course $\theta_{12}$ remains arbitrary in either cases. However the two types of symmetry
differ by their predictions on $\theta_{13}$ and the CP phase $\delta$. In case of $\mu-\tau$ 
permutation symmetry $\delta$ is washed out in the standard parametrization of PMNS matrix,
as a consequence of $\theta_{13}=0$. Thereby $\mu-\tau$ permutation symmetry naturally 
corresponds to CP conservation. In contrast to $\mu-\tau$ permutation symmetry, the reflection 
symmetry is featured with a non-zero $\theta_{13}$ 
and in addition, it corresponds to a maximal value of the CP phase ($\delta=\pm \pi/2$).
We can also note down that in the standard parametrization, if we restrict $\theta_{13}$
to be zero, $\mu-\tau$ reflection symmetry readily reproduces the properties of permutation
symmetry. In that sense $\mu-\tau$ reflection symmetry is a more general symmetry that 
can accommodate non-zero $\theta_{13}$ 
as well as CP violation. \\
\indent Bi-maximal (BM) mixing and tri-bimaximal (TBM) mixing are two special cases of 
$\mu-\tau$ permutation symmetry. It is obvious that these special mixing schemes or in general 
the permutation symmetric models are mostly explored in the phenomenological studies of lepton
mixing. In the present scenario, after the discovery of non zero $\theta_{13}$, permutation
symmetry is seemingly an inadequate theory as it can not accommodate non-zero $\theta_{13}$ 
and the CP violation. In this regard $\mu-\tau$ reflection symmetry might serve
a precious role in neutrino physics research. In comparison to the permutation symmetry, 
$\mu-\tau$ reflection symmetry and its possible phenomenological implications are less studied. 
The predictions $\theta_{23}=\pi/4$
and $\delta=\pm \pi/2$ are the central point of any $\mu-\tau$ reflection symmetric
model. Such models incorporated with non abelian discrete symmetries and their significance 
have been discussed in Refs.\cite{rnmoha,rnmoha1,nishi,he,ding,ema}. The implementation of
reflection symmetry in see-saw mechanism and the scenario of broken symmetry under
renormalization group running effects are studied in Refs.\cite{zhao,king1,nnath,nnath1}.  
Phenomenological consequences in other scenarios can be found in 
Refs.\cite{rode2,xing1,nishi1,zhao1,sgoswami}. \\ 
\indent Though $\mu-\tau$ reflection symmetry corresponds to the predictions- $\theta_{23}=\pi/4$
and $\delta=\pm \pi/2$, we can notice that the reflection symmetric nature of the mixing
matrix can not be viewed by direct substitution of these values in the lepton mixing matrix 
in standard parametrization. This is in contrast to $\mu-\tau$ permutation symmetry where
its predictions ($\theta_{23}=\pi/4$ and $\theta_{13}= 0$) directly results into a $\mu-\tau$ 
(permutation) symmetric mixing matrix upon substitution.
We have addressed this issue in this work and seek possible solutions to restore
the symmetry in the standard parametrization. A full parametrization of the lepton mixing
matrix with three mixing angles and six phases is considered and specific choice of the phases
is found to serve the purpose of this work. Besides the maximal Dirac phase $\delta$, as 
accommodated by reflection symmetry itself, we are led to additionally set maximal values
of Majorana phases ($\alpha$ and $\beta$) too in order to restore 
reflection symmetry property of the lepton 
mixing matrix. \\ 
\indent In view of the oscillation data, a small deviation 
of $\theta_{23}$ from its maximal value is also notable. A perturbation which can break the 
symmetry is necessary to account for the desired deviations. We consider the contributions 
from charged lepton sector as a possible scheme to deviate $\theta_{23}$ and $\delta$ from 
their maximal values. In a basis where charged lepton mass matrix is non-diagonal, the 
charged lepton mixing matrix is allowed to break the reflection symmetry of the neutrino
mixing matrix. The charged lepton mixing matrix can be parametrized in terms three mixing
angles and three complex phases. All of them act as free parameters in the present study.
On the neutrino sector atmospheric angle and the Dirac phase assume maximal values
through reflection symmetry while other two mixing angles remain unconstrained. The 
lepton mixing angles basically depend on the corresponding neutrino mixing angles
with small contributions from the charged lepton mixing parameters. It is the purpose
of this work to study in what way the various free parameters influence the predictions
of the lepton mixing angles. The correlation between the free parameters and the lepton
mixing angles are studied and possible values of the free parameters are examined
on the basis of numerical analysis.     \\
\indent The paper is organised as follow: in section 2 we outline the basic ingredients 
of lepton mixing which are necessary for the present study. In section 3 we briefly 
review $\mu-\tau$ reflection 
symmetry and discuss the ambiguity addressed. Section 4 discusses the scenario of broken 
reflection symmetry under the charged lepton correction scheme.  
Finally in section 5 we summarize and conclude the work.

\section{Ingredients of lepton mixing }
\indent Standard model charged current interaction Lagrangian for the leptons in flavour 
basis is given by
\begin{equation}
\mathcal{L}_{int} = -\frac{g}{\sqrt{2}} \bar{l^{\prime}_L} \gamma^{\mu} 
                                          \nu^{\prime}_L W^{-}_{\mu} + h.c.,
\end{equation}
 where $l^{\prime}_L=(e^{\prime} \ \mu^{\prime} \ \tau^{\prime})^T_L$ 
 and $\nu^{\prime}_L=(\nu^{\prime}_e \ \nu^{\prime}_{\mu} \ \nu^{\prime}_{\tau})^T_L$ represent
the left handed charged lepton flavour states and neutrino flavour states respectively. In 
transforming to mass basis we get the lepton mixing $U$, also known as the PMNS matrix, in the 
Lagrangian :  
\begin{equation}
\mathcal{L}_{int} = -\frac{g}{\sqrt{2}} \bar{l_L} \gamma^{\mu} U \nu_L W^{-}_{\mu} + h.c..
\end{equation}
The un-primed fields, viz. $l_L=(e \ \mu \ \tau)^T_L$ and $\nu_L=(\nu_1 \ \nu_2 \ \nu_3)^T_L$, 
denote the respective mass eigenstates. We define the diagonalizing matrices $U_l$ and $U_{\nu}$
for the charged lepton and Majorana neutrino mass matrices respectively as : 
$U^{\dagger}_lM^{\dagger}_lM_lU_l=M^2_{ld} \equiv Diag(m^2_e, m^2_{\mu}, m^2_{\tau})$
and $U^{\dagger}_{\nu}M_{\nu}U^{*}_{\nu}=M_{\nu d} \equiv Diag(m_1, m_2, m_3)$, 
such that the PMNS matrix is given by
\begin{equation}
U= U^{\dagger}_lU_{\nu}.
\end{equation}
If we choose the basis where flavour eigenstates and mass eigenstates of the 
charged leptons are identical, the charged lepton mixing matrix $U_l$ in Eq.(3) becomes
an identity matrix. The PMNS matrix $U$, which is a unitary matrix, can be parametrized in 
terms three mixing angles and six phases. In a familiar parametrization $U$ can be expressed 
as
\begin{equation}
U = P_1 V P_2,
\end{equation}
where the mixing matrix $V$ is parametrized in terms of the three mixing angles 
($\theta_{12}$, $\theta_{12}$, $\theta_{12}$) and the Dirac CP phase $\delta$. The diagonal
phase matrix  $P_2=diag( e^{i \alpha}, e^{i \beta},1)$ contains two Majorana phases $\alpha$ 
and $\beta$ while $P_1=diag( e^{i \phi_1}, e^{i \phi_2},e^{i \phi_3})$ contains the remaining
three phases. The three phases in $P_1$ are un-physical which can be eliminated from the mixing 
matrix $U$ by phase redefinition of the charged lepton fields. In the standard parametrization 
we have
\begin{equation}
       V = \begin{pmatrix}
    c_{12} c_{13}                       & s_{12} c_{13} 
                                                          & s_{13} e^{-i \delta}\\
    -s_{12} c_{23}-c_{12} s_{23} s_{13}e^{i \delta} & c_{12} c_{23}-s_{12} s_{23} s_{13} e^{i \delta}
                                                          & s_{23} c_{13}\\
    s_{12} s_{23}-c_{12} c_{23} s_{13}e^{i \delta} & -c_{12} s_{23}-s_{12} c_{23} s_{13} e^{i \delta} 
                                                          & c_{23} c_{13} \\ 
                                                      \end{pmatrix}, 
\end{equation}
where $s_{ij}=\sin \theta_{ij}$ and $c_{ij}=\cos \theta_{ij}$ with $ij=12,23,13$. So far physical
observables are concerned, one may simply drop $P_1$ from Eq.(4). If neutrinos are considered 
as Dirac particles $P_2$ can further be dropped in a particular study.  \\
\begin{table}[]
\centering
 \begin{tabular}{ccc}
\hline
       Mixing Parameter & Best fit & 3 $\sigma$  \\
\hline
            $\sin^2\theta_{12}$ & $0.310$ & $0.275$ - $0.350$ \\
          $\sin^2\theta_{23}$ & $0.580$ & $0.418$ - $0.627$    \\
          $\sin^2\theta_{13}$ & $0.0224$ & $0.0204$ - $0.0244$  \\
          $\delta$ & $215^{\circ}$ & $125^{\circ}$ - $392^{\circ}$ \\
\hline              
\end{tabular}
\caption{Best fit and $3\sigma$ values of mixing parameters
 for normal hierarchy (NH) from global analysis \cite{global1}.}    
\end{table}
\indent Turning to the mixing angles, the sine of the angles can be expressed in terms 
of the absolute values of the elements of $U$ as follows :
\begin{equation}
\sin^2\theta_{13} = \abs{U_{e3}}^2, \ \ \ \sin^2\theta_{12}= \frac{\abs{U_{e2}}^2}{1-\abs{U_{e3}}^2},
                    \ \ \ \sin^2\theta_{23}= \frac{\abs{U_{\mu 3}}^2}{1-\abs{U_{e3}}^2}.
\end{equation}
The measure of CP violation is expressed in terms of parametrization independent quantities
called rephasing invariants. We consider the Jarlskog invariant \cite{cjarl} given by
\begin{equation}
J= Im[U_{e2}U_{\mu 3}U^{*}_{e3}U^{*}_{\mu 2}],
\end{equation}
for our analysis. For the mixing matrix $V$ in Eq.(5), Eq.(7) yields
\begin{equation}
J= s_{12}c_{12}s_{23}c_{23}s_{13}c^2_{13}\sin\delta.
\end{equation}
The best fit and $3\sigma$ values of the three mixing angles and the Dirac CP phase 
 for normal hierarchy (NH) are presented in Table 1 from the global analysis \cite{global1}.

\section{$\mu-\tau$ reflection symmetry}

\indent To begin with, we first consider the flavor basis where charged lepton mass matrix
is diagonal such that there is no contribution to the PMNS matrix $U$ in Eq.(3) from
the charged lepton mixing matrix $U_l$. The discussion made in this section will be
concerned with the PMNS matrix without any charged lepton correction. In the next section
we will turn into the flavor basis with a non diagonal charged lepton mass matrix and
its effects on the PMNS matrix will be studied. \\  
\indent  The original formulation of $\mu-\tau$ reflection symmetry, introduced by Harrison and 
Scott \cite{hs}, concerns the Dirac phase $\delta$ only where Majorana phases are dropped
from the lepton mixing matrix. In the present consideration it is represented by the 
mixing matrix $V$ in Eq.(5). They were motivated from the observation that the modulus 
of each $\mu$-flavor element of the mixing matrix is approximately equal to that of the 
corresponding $\tau$-flavor 
element (i.e. $\abs{V_{\mu i}} \simeq \abs{V_{\tau i}}$),
as revealed by the neutrino oscillation data. They follow
a specific parametrization of the mixing matrix based on the assumption  
$\abs{U_{\mu i}} = \abs{U_{\tau i}}$, and arrive at the mixing matrix
\begin{equation}
V_{HS} = \begin{pmatrix}
            u_1 & u_2  & u_3  \\
             v_1 & v_2  & v_3  \\
              v^{*}_1 & v^{*}_2  & v^{*}_3  \\
                \end{pmatrix},
\end{equation}
where $u_i$'s are taken as real and $v_i$'s as complex. This mixing matrix is
symmetric under a combined operation of interchanging $\nu_{\mu}$ and $\nu_{\tau}$ 
flavour states and complex conjugation of the mixing matrix. This combined operation
of symmetry is referred to as $\mu-\tau$ reflection symmetry. The corresponding mass 
matrix is required to invariant under the $\mu-\tau$ reflection operation which can be 
expressed as
\begin{equation}
\left( A_{\mu\tau} M A_{\mu\tau} \right)^{*} = M,
\end{equation}
where 
\begin{equation}
A_{\mu\tau} = \begin{pmatrix}
            1 &  0  &  0  \\
            0 &  0  &  1 \\
            0 &  1  &  0  \\
                \end{pmatrix},
\end{equation}
is the $\mu-\tau$ exchange operator. The mass matrix satisfying Eq.(10) is given by 
\begin{equation}
M = \begin{pmatrix}
            M_{ee} &  M_{e\mu}  &  M^{*}_{e\mu}  \\
            M_{e\mu} &  M_{\mu\mu}  &  M_{\mu\tau} \\
            M^{*}_{e\mu} &  M_{\mu\tau} &  M^{*}_{\mu\mu} \\
                \end{pmatrix},
\end{equation}
where the elements $M_{ee}$ and  $M_{\mu\tau}$ are real. This mass matrix were reproduced 
in an $A_4$ based model by Babu, Ma and Valle, one year after the concept of reflection 
symmetry introduced.
The crucial thing about the mixing matrix $V_{HS}$ is that it is linked with the aforementioned
predictions- $\theta_{23}=\pi/4$ and $\delta= \pm \pi/2$. To see this connection let us 
consider the Jarlskog's invariant for $V_{HS}$ which is given by 
$J= \frac{1}{2}u_1 u_2 u_3$, as obtained from Ref.\cite{hs}. In terms of mixing matrix 
elements modulus of $J$ can be written as
\begin{equation}
\abs{J} = \frac{1}{2} \abs{V_{e1} V_{e2} V_{e3}}.
\end{equation}
Again in the standard parametrization, from Eq.(8) we have
\begin{equation}
\abs{J} = \frac{1}{2} \abs{V_{e1} V_{e2} V_{e3} }\abs{\sin\delta} \sin 2\theta_{23}.
\end{equation}
For non zero $\theta_{13}$, comparison of Eqs.(13) and (14) gives
$\abs{\sin\delta} \sin 2\theta_{23}=1$. For $\theta_{23}=\pi/4$, this implies
$\delta= \pm \pi/2$.  \\
\indent Conversely we may now wish to see whether the mixing matrix $V$ in Eq.(5), with 
$\theta_{23}=\pi/4$ and $\delta= \pm \pi/2$, reflect the reflection symmetric nature
of $V_{HS}$ or not. To be explicit, we have
\begin{equation}
       V = \begin{pmatrix}
    c_{12} c_{13}    & s_{12} c_{13}  & \mp i s_{13} \\
         \frac{1}{\sqrt{2}}(-s_{12} \mp i c_{12} s_{13}) 
                & \frac{1}{\sqrt{2}} (c_{12} \mp i s_{12} s_{13}  & \frac{c_{13}}{\sqrt{2}}  \\
            \frac{1}{\sqrt{2}} (s_{12} \mp i c_{12}s_{13}
                &  \frac{1}{\sqrt{2}}(-c_{12} \mp i s_{12} s_{13} 
                                                          & \frac{c_{13}}{\sqrt{2}} \\ 
                                                      \end{pmatrix}, 
\end{equation}
from Eq.(5), where '$\mp$' sign corresponds to $\delta= \pm \pi/2$. We can see that the
matrix reproduces the presumed conditions : $\abs{U_{\mu i}} = \abs{U_{\tau i}}$, as expected.
However $\mu$ and $\tau$-flavour mixing elements of $V$ do not satisfy 
$V_{\tau j}= V^{*}_{\mu j}$ (followed from Eq.(9)), instead we have 
$V_{\tau j}= -V^{*}_{\mu j}$ for $j=1,2$ while $V_{\tau 3}=V^{*}_{\mu 3}$. Further, most 
significantly, first row elements of the mixing matrix are not real which necessarily 
violates the inherent reflection symmetric nature carried by $V_{HS}$.
That means the PMNS matrix in the standard parametrization does not exhibit reflection 
symmetry under the constraints $\theta_{23}=\pi/4$ and $\delta= \pm \pi/2$. 
It is however necessary to point out that the mass matrix 
diagonalized by $V$ in Eq.(15) respects reflection
symmetry (Eq.(12)).\\
\indent To realize the properties of reflection symmetry in the 'standard' PMNS matrix 
consistent with $\theta_{23}=\pi/4$ and $\delta= \pm \pi/2$, we find it useful to consider 
the full parametrization defined in Eq.(4). All the six phases taken into account, we get 
the PMNS matrix from Eq.(4) as
\begin{equation}
 U = \begin{pmatrix}
       V_{e1} e^{i(\alpha + \phi_1)}    & V_{e2} e^{i(\beta + \phi_1)}  & V_{e3} e^{i\phi_1} \\
         V_{\mu 1} e^{i(\alpha + \phi_2)}    & V_{\mu 2} e^{i(\beta + \phi_2)}  & V_{\mu 3} e^{i\phi_2} \\
          V_{\tau 1} e^{i(\alpha + \phi_3)}    & V_{\tau 2} e^{i(\beta + \phi_3)}  & V_{\tau 3} e^{i\phi_3} \\
                                                      \end{pmatrix},
\end{equation}
with the elements $V_{lj}$ ($l=e,\mu,\tau$; $j=1,2,3$) defined through Eq.(5). Compared 
to the 'Dirac like' mixing matrix $V$ (Eq.(5)), concerned with the original formulation 
of reflection symmetry, we now have five additional phases under consideration. With 
$\theta_{23}=\pi/4$ and $\delta= \pm \pi/2$ the elements $V_{lj}$ are given in Eq.(15). 
We find that the reflection symmetric nature of $V_{HS}$ can be restored in $U$ with 
proper choice of these additional phases. Let us first consider the case $\delta= \pi/2$. 
We then conveniently choose $\phi_1=\pi/2$ and $\alpha =\beta = -(\pi/2)$ 
to make the first row elements all real. In addition the remaining two phases are 
constrained to zero ($\phi_2=\phi_3=0$). With these specific values of the phases the PMNS
matrix in Eq.(16) becomes
\begin{equation}
       U = \begin{pmatrix}
    c_{12} c_{13}    & s_{12} c_{13}  &  s_{13} \\
         \frac{1}{\sqrt{2}}s(-c_{12}s_{13}) + is_{12})
                & \frac{1}{\sqrt{2}} (-s_{12} s_{13} -ic_{12})  & \frac{c_{13}}{\sqrt{2}}  \\
            \frac{1}{\sqrt{2}}(-c_{12}s_{13}) - is_{12})
                &  \frac{1}{\sqrt{2}} (-s_{12} s_{13} +ic_{12})
                                                          & \frac{c_{13}}{\sqrt{2}} \\ 
                                                      \end{pmatrix}. 
\end{equation}
This matrix is now exactly similar to $V_{HS}$ with the first row elements all real and 
second and third row elements satisfying the condition
$U_{\tau j}= U^{*}_{\mu j}$ for all $j=1,2,3$. The specific values of the un-physical 
phases so chosen may be attributed to the arbitrariness in their values. Besides, the 
values of the Majorana phases are remarkable. It is meant that, in addition to maximal 
$\delta$, Majorana phases are also enforced to pick up maximal values in order to 
restore the symmetry. For the other case with $\delta= -\pi/2$, we may have the choices: 
$\phi_1= -\pi/2$ and $\alpha =\beta = (\pi/2)$, which differ by a negative sign in 
comparison to the previous set of values. The other two phases $\phi_2$ and $\phi_3$ 
should be kept fixed at zero as before. The resulting PMNS matrix, containing the 
reflection symmetry, is similar to that in Eq.(17) but with the elements $U_{\mu 1}$ 
and $U_{\mu 2}$ replaced by complex conjugation of the respective elements of $U$ in 
Eq.(17). In other sense the two mixing matrices are related by
\begin{equation}
U_{\delta=-\pi/2} = U^{*}_{\delta=\pi/2},
\end{equation}
where $U_{\delta=\pi/2}$ represents the matrix in Eq.(17).\\
\indent In the basis where charged lepton mass matrix is already diagonal, the Majorana 
neutrino mass matrix can be obtained from $M=U M_{\nu d}U^{T}$. The mixing matrix in 
Eq.(17) leads to a mass matrix satisfying the reflection symmetry as shown by $M$ in Eq.(12). 
The elements are given by
\begin{align}
M_{ee}= & \left( m_1c^2_{12} + m_2s^2_{12} \right) c^2_{13} +m_3s^2_{13}, \\ \nonumber
M_{\mu\tau}= & \frac{1}{2}m_1\left( c^2_{12}s^2_{13} + s^2_{12}\right)
                + \frac{1}{2}m_2\left( s^2_{12}s^2_{13} + c^2_{12}\right) +\frac{1}{2}m_3c^2_{13}, \\ \nonumber  
M_{e\mu}= & \frac{1}{\sqrt{2}}\left( -m_1c^2_{12} - m_2s^2_{12} +m_3\right)s_{13}c_{13}
                            +\frac{i}{\sqrt{2}}(m_1-m_2)s_{12}c_{12}c_{13}, \\ \nonumber
M_{\mu\mu}= & \frac{1}{2} \left[ m_1\left( c^2_{12}s^2_{13}-s^2_{12} \right)
                                  + m_2\left( s^2_{12}s^2_{13}-c^2_{12} \right) + m_3c^2_{13} \right] 
                                                  -i (m_1-m_2)s_{12}c_{12}s_{13}.
\end{align} 
For the case $\theta_{23}=\pi/4$ and $\delta= -\pi/2$, the mass matrix obtained from 
$U_{\delta=-\pi/2}$ follows a similar connection as presented in Eq.(18). The mass matrices
of the two cases are complex conjugate of each other 
($M_{\delta=-\pi/2} = M^{*}_{\delta=\pi/2}$).

\section{Charged lepton contributions}

\indent If the values of $\theta_{23}$ and $\delta$ are not exactly maximal, one has to 
deviate from the reflection symmetry in some way. Possible corrections from the charged 
lepton sector are often considered in this regard 
\cite{hosmuth,rode,rode1,king,antusch,marzoca,marzoca1,duarah,sroy,sdev,sdev1,sroy1}. 
To employ the charged lepton correction scheme we recall Eq.(3) and define the lepton 
mixing matrix as 
\begin{equation}
\tilde{U}^{MNS} = (\tilde{U}^l)^{\dagger}U^{\nu},
\end{equation}
in the basis where charged lepton mass matrix is non-diagonal. Under this basis 
the lepton mixing matrix will contain contributions from both $\tilde{U}^l$ and 
$U^{\nu}$.  The common idea of this approach is to assume a perfect symmetry in 
either of the two sectors ($\tilde{U}^l$ or $U^{\nu}$) and then perturb this 
symmetry by the other leading to a desired lepton mixing matrix. A treatment involving both 
the alternate cases is available in Ref.\cite{hosmuth}. The symmetry considered in most 
works is the $\mu-\tau$ permutation symmetry which incorporates maximal atmospheric angle 
and zero reactor angle while solar angle is left arbitrary. Since Bimaximal mixing and 
Tri-bimaximal mixing are two special cases of $\mu-\tau$ permutation symmetry, deviations
from such special mixing through charged lepton correction is most common. However 
corrections to special mixing based on $\mu-\tau$ reflection symmetry from 
charged lepton sector is very rare in the
literature. \\
\indent Each of $\tilde{U}^l$ and $U^{\nu}$ is a unitary matrix and can be parametrized in terms 
of three mixing angles and six phases as well. We invoke the parametrization set up presented
in Eq.(4) to define both the mixing matrices and we get
\begin{equation}
\tilde{U}^l= P^l_1 V^l P^l_2, \ \ \ \ \ \ U^{\nu}= P^{\nu}_1 V^{\nu} P^{\nu}_2.
\end{equation}
With the above matrices the resulting PMNS matrix in Eq.(20) becomes
\begin{equation}
\tilde{U}^{MNS}= (P^l_2)^{\dagger}(V^l)^{\dagger}(P^l_1)^{\dagger}P^{\nu}_1 V^{\nu} P^{\nu}_2.
\end{equation}
The diagonal phase matrices are defined as :
 $P^l_1=diag( e^{i \phi^l_1}, e^{i \phi^l_2},e^{i \phi^l_3})$,
 $P^l_2=diag( e^{i \alpha^l}, e^{i \beta^l},1)$,
 $P^{\nu}_1=diag( e^{i \phi^{\nu}_1}, e^{i \phi^{\nu}_2},e^{i \phi^{\nu}_3})$,
 $P^{\nu}_2=diag( e^{i \alpha^{\nu}}, e^{i \beta^{\nu}},1)$; while the matrices $V^l$ and 
 $V^{\nu}$ resemble $V$ in Eq.(5), given by
\begin{equation}
       V^l = \begin{pmatrix}
                  c^l_{12} c^l_{13}   & s^l_{12} c^l_{13} 
                                                          & s^l_{13} e^{-i \delta^l}\\
    -s^l_{12} c^l_{23}-c^l_{12} s^l_{23} s^l_{13}e^{i \delta^l} 
                     & c^l_{12} c^l_{23}-s^l_{12} s^l_{23} s^l_{13} e^{i \delta^l}
                                                          & s^l_{23} c^l_{13}\\
      s^l_{12} s^l_{23}-c^l_{12} c^l_{23} s^l_{13}e^{i \delta^l} 
                     & -c^l_{12} s^l_{23}-s^l_{12} c^l_{23} s^l_{13} e^{i \delta^l} 
                                                          & c^l_{23} c^l_{13} \\ 
                                                      \end{pmatrix},
\end{equation}
\begin{equation}
       V^{\nu} = \begin{pmatrix}
    c^{\nu}_{12} c^{\nu}_{13}   & s^{\nu}_{12} c^{\nu}_{13} 
                                              & s^{\nu}_{13} e^{-i \delta^{\nu}}\\
    -s^{\nu}_{12} c^{\nu}_{23}-c^{\nu}_{12} s^{\nu}_{23} s^{\nu}_{13}e^{i \delta^{\nu}} 
                     & c^{\nu}_{12} c^{\nu}_{23}-s^{\nu}_{12} s^{\nu}_{23} s^{\nu}_{13} e^{i \delta^{\nu}}
                                                          & s^{\nu}_{23} c^{\nu}_{13}\\
      s^{\nu}_{12} s^{\nu}_{23}-c^{\nu}_{12} c^{\nu}_{23} s^{\nu}_{13}e^{i \delta^{\nu}} 
                     & -c^{\nu}_{12} s^{\nu}_{23}-s^{\nu}_{12} c^{\nu}_{23} s^{\nu}_{13} e^{i \delta^{\nu}} 
                                                          & c^{\nu}_{23} c^{\nu}_{13} \\ 
                                                      \end{pmatrix}.
\end{equation}
To take into account charged lepton contributions we now assume the exact 
$\mu-\tau$ reflection symmetry, presented in section 3, in the neutrino sector 
and let this symmetry perturb by the the charged lepton mixing matrix 
$\tilde{U}^{l}$. We set the specific values of the neutrino mixing parameters
as : $\theta^{\nu}_{23}=\pi/4$, $\delta^{\nu}= \pm \pi/2$, 
$\alpha^{\nu}=\beta^{\nu}=\mp \pi/2$, $\phi^{\nu}_1=\pm \pi/2$ and 
$\phi^{\nu}_2=\phi^{\nu}_3=0$. With these substitutions we get the reflection
symmetric neutrino mixing matrix given by
\begin{equation}
U^{\nu} = \begin{pmatrix}
    c^{\nu}_{12} c^{\nu}_{13}    & s^{\nu}_{12} c^{\nu}_{13}  &  s^{\nu}_{13} \\
         \frac{1}{\sqrt{2}}\left( -c^{\nu}_{12}s^{\nu}_{13} \pm is^{\nu}_{12}\right) 
                 & \frac{1}{\sqrt{2}}\left(-s^{\nu}_{12} s^{\nu}_{13} \mp ic^{\nu}_{12}\right)   
                                             & \frac{1}{\sqrt{2}}c^{\nu}_{13}  \\
            \frac{1}{\sqrt{2}} \left( -c^{\nu}_{12}s^{\nu}_{13} \mp is^{\nu}_{12}\right) 
                                &  \frac{1}{\sqrt{2}} \left(-s^{\nu}_{12} s^{\nu}_{13} \pm ic^{\nu}_{12}\right)
                                                          & \frac{1}{\sqrt{2}}c^{\nu}_{13} \\ 
                                                      \end{pmatrix}, 
\end{equation}
in Eq.(22). The '$\pm$' signs in the '$21$'-element of the above mixing matrix
 correspond to the two different cases- viz.,
{\it Case I}: $\theta^{\nu}_{23}=\pi/4$, $\delta^{\nu}= + \pi/2$;
and {\it Case II}: $\theta^{\nu}_{23}=\pi/4$, $\delta^{\nu}= - \pi/2$; 
respectively.                  \\
\indent The mixing angles $\theta^{\nu}_{12}$ and $\theta^{\nu}_{13}$ in $U^{\nu}$
in Eq.(25) are left arbitrary by the reflection symmetry. After having this reflection 
symmetric neutrino mixing matrix, we are left with the charged lepton mixing matrix 
$\tilde{U}^{l}$ in Eq.(22). All the mixing parameters in $\tilde{U}^{l}$, viz., 
three mixing angles- $\theta^{l}_{12}$, $\theta^{l}_{23}$, $\theta^{l}_{13}$,
and six phases- $\delta^l$, $\alpha^l$, $\beta^l$, $\phi^l_1$, $\phi^l_2$, $\phi^l_3$,
remain as free parameters in the model. In total we have eleven free parameters 
that determine the mixing parameters of the PMNS matrix $\tilde{U}^{MNS}$ in Eq.(22).
However we can identify three of the six phases in $\tilde{U}^{l}$ as un-physical
and the total number of free parameters can be reduced. 
Note that $\tilde{U}^{MNS}$ is also parametrized in the same manner 
as $\tilde{U}^l$ and $U^{\nu}$ which is given by 
\begin{equation}
\tilde{U}^{MNS} = P^{MNS}_1 V^{MNS} P^{MNS}_2,
\end{equation}
with $P^{MNS}_1=diag( e^{i \phi_1}, e^{i \phi_2},e^{i \phi_3})$, 
$P^{MNS}_2=diag( e^{i \alpha}, e^{i \beta},1)$ and $V^{MNS}$ given by Eq.(5). This 
PMNS matrix contains total six phases out of which the three phases in the phase matrix 
$P^{MNS}_1$ are un-physical. Again from Eq.(22) we see that five out of the six phases
of $\tilde{U}^l$ are distributed in the phase matrices $P^l_1$ and $P^l_2$. We can 
commute $P^l_1$ to the left of the right hand side of Eq.(22) and it helps us to 
identify the un-physical phases present in Eq.(22). By doing so, Eq.(22) can be 
re-expressed as
\begin{equation}
\tilde{U}^{MNS}= (P^l)^{\dagger}(U^l)^{\dagger} U^{\nu},
\end{equation}   
where $(P^l)^{\dagger}=(P^l_2)^{\dagger}(P^l_1)^{\dagger}$ and $(U^l)^{\dagger}$ is given
by
\begin{equation}
\begin{pmatrix}
c^l_{12}c^l_{13} & s^l_{12}c^l_{23}e^{i\delta^l_{12}} 
                  -c^l_{12}s^l_{23}s^l_{13}e^{-i(\delta^l_{23}-\delta^l_{13})} 
                          & s^l_{12}s^l_{23}e^{i(\delta^l_{12}+\delta^l_{23})} 
                               -c^l_{12}c^l_{23}s^l_{13}e^{i\delta^l_{13}}  \\
s^l_{12}c^l_{13}e^{-i\delta^l_{12}}  & c^l_{12}c^l_{23}
                   -s^l_{12}s^l_{23}s^l_{13}e^{-i(\delta^l_{12}+\delta^l_{23}-\delta^l_{13})} 
                               & -c^l_{12}s^l_{23}e^{i\delta^l_{23})} 
                                 -s^l_{12}c^l_{23}s^l_{13}e^{-i(\delta^l_{12}-\delta^l_{13})}  \\ 
 s^l_{13}e^{-i\delta^l_{13}}  &  s^l_{23}c^l_{13}e^{-i\delta^l_{23}}  &  c^l_{23}c^l_{13}  \\                                                                
\end{pmatrix}.
\end{equation}
Above charged lepton mixing matrix is derived from the relation 
$(U^l)^{\dagger} = (U^l_{12})^{\dagger}(U^l_{13})^{\dagger}(U^l_{23})^{\dagger}$,
where the form of the rotation matrices $U^l_{12}$, $U^l_{13}$ and $U^l_{23}$ are
presented in appendix A with each mixing parameter symbolized with the superscript
$'l'$. The phases newly defined in $U^{l}$ satisfy the relations
\begin{equation}
\delta^l_{12} = \phi^l_1 - \phi^l_2, \ \ \ \delta^l_{13} =\delta^l- \phi^l_1+ \phi^l_3, 
               \ \ \ \delta^l_{23} = \phi^l_2 - \phi^l_3.
\end{equation}  
In view of Eqs.(26) and (27), we can now drop the phase matrices $P^{MNS}_1$ and 
$(P^l)^{\dagger}$ from these equations as they stand for the un-physical phases. 
The resulting PMNS matrix is given by
\begin{equation}
V^{MNS} P^{MNS}_2=U^{MNS}= (U^l)^{\dagger} U^{\nu},
\end{equation}
with $U^l$ and $U^{\nu}$ given in Eqs.(28) and (25) respectively.
The PMNS matrix in Eq.(30), now contains the Dirac phase $\delta$ and the Majorana 
phases $\alpha$ and $\beta$ only on the left hand side. After the elimination of 
the un-physical phases, number of free parameters on the right hand side of Eq.(30)
reduces to eight with three complex phases from the charged lepton sector. The
mixing angles and the CP phases of $U^{MNS}$ will be the functions of these eight 
free parameters. \\
\indent Naturally it is expected that the lepton mixing angles $\theta_{12}$, 
$\theta_{23}$ and $\theta_{13}$ receive their major contributions from the respective 
neutrino mixing angles ($\theta^{\nu}_{12}$, $\theta^{\nu}_{23}$, $\theta^{\nu}_{13}$) 
with small contributions from the charged lepton sector.
By the recent discovery third angle $\theta_{13}$ is found to be very small compared
to the other two. By the same token we can expect $\theta^{\nu}_{13}$ 
($\sim \theta_{13}$), to be very small compared to $\theta^{\nu}_{12}$ and  $\theta^{\nu}_{23}$. 
To analyze the consequences of Eq.(30) we are interested to consider a mixing matrix
$U^{l}$ with small mixing angles $\theta^{l}_{12}$, $\theta^{l}_{23}$ and 
$\theta^{l}_{13}$ \cite{hosmuth,king}. Under this consideration it is useful to 
define the parameters $\sin\theta^{l}_{ij}=\lambda_{ij} >0$ with $ij=12,\ 23,\ 13$.
As $\theta^{\nu}_{13}$ should be smaller than $\theta_{13}$, we can assume an upper
bound $\sim 0.15$ for $\sin\theta^{\nu}_{13}$ as per the best fit value of 
$\sin\theta_{13}$ \cite{global1}. Under such situation we also have 
$O(s^{\nu}_{13})\sim O(\lambda_{ij})$ for the small $\lambda_{ij}$'s.
With these parameters mixing matrix $U^l$ in Eq.(28) becomes
\begin{equation}
(U^l)^{\dagger} \approx
\begin{pmatrix}
      1-\frac{1}{2}\lambda^2_{12}-\frac{1}{2}\lambda^2_{13}
              & -\lambda_{12}e^{i\delta^l_{12}} -\lambda_{23}\lambda_{13}e^{-i(\delta^l_{23}-\delta^l_{13})}
              & -\lambda_{13}e^{i\delta^l_{13}} -\lambda_{12}\lambda_{23}e^{i(\delta^l_{12}+\delta^l_{23})} \\
     \lambda_{12}e^{-i\delta^l_{12}} & 1-\frac{1}{2}\lambda^2_{12}-\frac{1}{2}\lambda^2_{23}   
              & -\lambda_{23}e^{i\delta^l_{23}} -\lambda_{12}\lambda_{13}e^{-i(\delta^l_{12}-\delta^l_{13})} \\ 
     \lambda_{13}e^{-i\delta^l_{13}} & \lambda_{23}e^{-i\delta^l_{23}}
              &  1-\frac{1}{2}\lambda^2_{23}-\frac{1}{2}\lambda^2_{13}  \\  
\end{pmatrix},
\end{equation}
where higher order terms with $\lambda^n_{ij}$ ($n>2$) are neglected for the small
angles. The elements of $U^{MNS}$ are then given by Eqs.(30), (31) and (25). For 
convenience we write down the following three elements:  
\begin{align}
U^{MNS}_{e3} & \simeq \sin\theta^{\nu}_{13} -\frac{1}{\sqrt{2}}
                       \left( \lambda_{12}e^{i\delta^l_{12}}+ \lambda_{13}e^{i\delta^l_{13}} \right) 
                                                                      \\  \nonumber
             & \ \ \ \ \ \ \ \ \ \ \ \ +\frac{1}{\sqrt{2}} 
                           \left( \lambda_{12}\lambda_{23}e^{i(\delta^l_{12}+\delta^l_{23})}
                            -\lambda_{23}\lambda_{13}e^{-i(\delta^l_{23}-\delta^l_{13})} \right),
\end{align}
\begin{figure}[]
\centering
\subfloat[]{\includegraphics[scale=0.42]{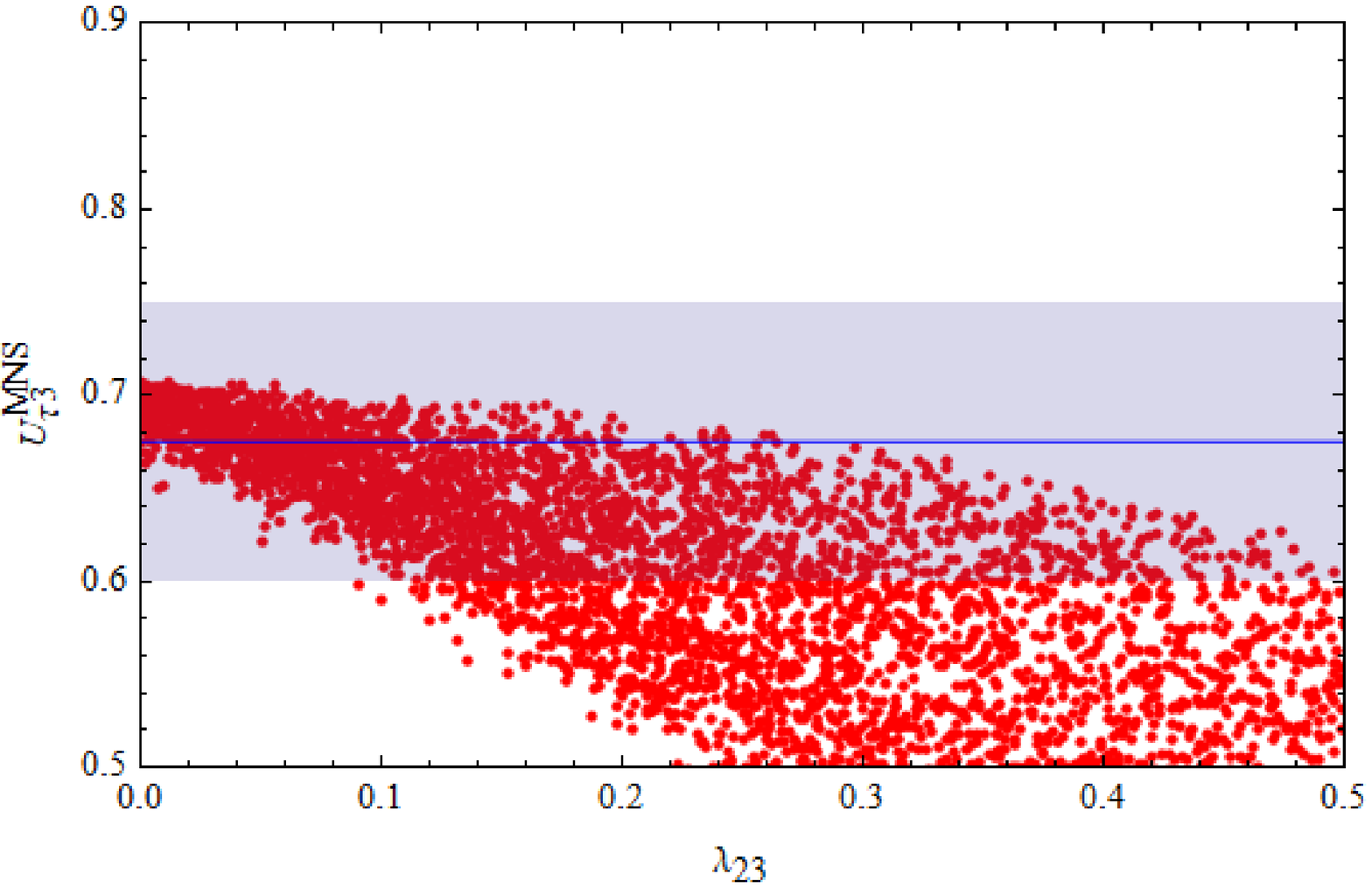}}
\subfloat[]{\includegraphics[scale=0.42]{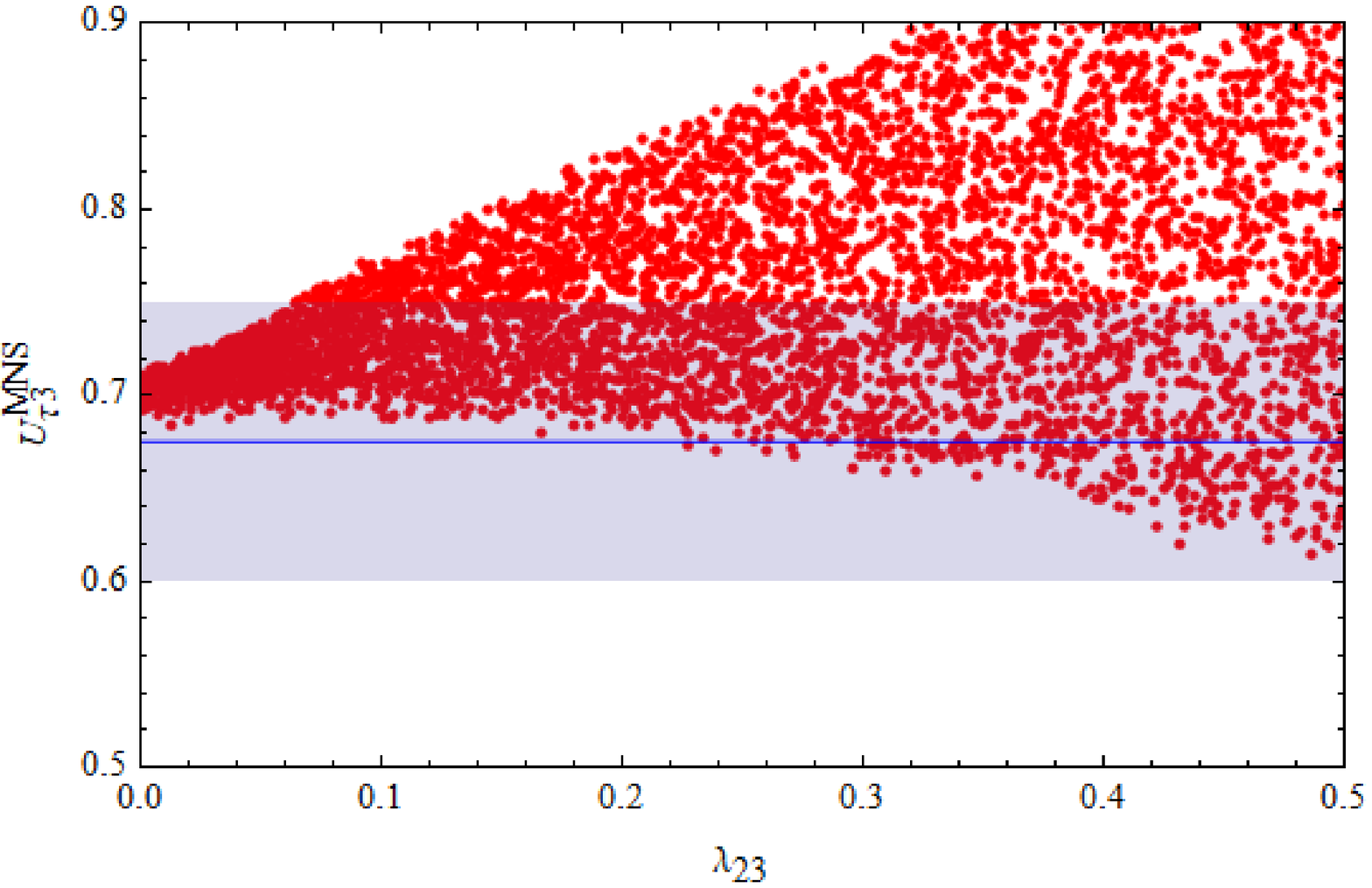}}
\caption{Correlation between $U^{MNS}_{\tau 3}$ and $\lambda_{23}$ for (a) negative 
values of $\cos\delta^l_{23}$ and (b) positive values of $\cos\delta^l_{23}$. 
Horizontal coloured bands represent the $3\sigma$ range and the blue lines stand 
for the best fit value of $U^{MNS}_{\tau 3}$ obtained from global data.}
\end{figure}
\begin{align}
U^{MNS}_{\mu 3} & \simeq  \frac{1}{\sqrt{2}} -\frac{1}{\sqrt{2}} \lambda_{23}e^{i\delta^l_{23}}
                           +\sin\theta^{\nu}_{13} \lambda_{12}e^{-i\delta^l_{12}}  \\  \nonumber
                & \ \ \ \ \ \ \ \ \ \ \ \  -\frac{1}{2\sqrt{2}}\left( \lambda^2_{12}+\lambda^2_{23}
                       +\sin^2\theta^{\nu}_{13} +2\lambda_{12}\lambda_{13}e^{-i(\delta^l_{12}-\delta^l_{13})} 
                                                                \right), 
\end{align} 
\begin{equation}
U^{MNS}_{\tau 3} \simeq \frac{1}{\sqrt{2}} +\frac{1}{\sqrt{2}}\lambda_{23}e^{-i\delta^l_{23}}
                           +\sin\theta^{\nu}_{13}\lambda_{13}e^{-i\delta^l_{13}} 
                            -\frac{1}{2\sqrt{2}}\left( \lambda^2_{23}+\lambda^2_{13}
                              +\sin^2\theta^{\nu}_{13} \right),  
\end{equation}
It can be noted from the left side of Eq.(30) that '$\mu 3$' and '$\tau 3$' elements 
of the PMNS matrix are free from the Majorana CP phases and are also real. Comparing 
the real and imaginary parts of Eq.(34) we get
\begin{equation}
\cos\theta_{23}\cos\theta_{13} = \frac{1}{\sqrt{2}} + \frac{1}{\sqrt{2}} \lambda_{23}\cos\delta^l_{23}
                                  + s^{\nu}_{13} \lambda_{13}\cos\delta^l_{13}
                   -\frac{1}{2\sqrt{2}}\left( \lambda^2_{23} +\lambda^2_{13}+(s^{\nu}_{13})^2\right),                        
\end{equation}
\begin{equation}
\sin\theta^{\nu}_{13}= -\frac{1}{\sqrt{2}}
                \frac{\lambda_{23}\sin\delta^l_{23}}{\lambda_{13}\sin\delta^l_{13}},
\end{equation}
and the imaginary parts of Eq.(33) give
\begin{equation}
\sin\theta^{\nu}_{13}= -\frac{1}{\sqrt{2}}\frac{\lambda_{23}\sin\delta^l_{23}
                        -\lambda_{12}\lambda_{13}\sin(\delta^l_{12}-\delta^l_{13})}
                          {\lambda_{12}\sin\delta^l_{12}}.
\end{equation}
Expressions (36) and (37) show the interconnection between the free parameters of $U^l$
and $\sin\theta^{\nu}_{13}$. These two equations can be used to constrain any two of 
the free parameters.
The factor $1/\sqrt{2}$ in Eq.(35) carries the sign of maximal atmospheric 
mixing while the second term contributes as first order in $\lambda_{ij}$. 
Rest of the terms in this expression account for the second order 
contributions which are relatively very small. We first study the correlation 
between $U^{MNS}_{\tau 3}$ and $\lambda_{23}$ from Eq.(35) with $0<s^{\nu}_{13}\leq 0.15$
and an approximate bound of $0-0.25$ for $\lambda_{13}$. We find that $\cos\delta^l_{23}$ 
plays a significant role in predicting $U^{MNS}_{\tau 3}$ within the observed bound
of global data. We inspect the correlation separately for positive values
($0\leq \cos\delta^l_{23} \leq 1$) as well as negative values ($-1\leq \cos\delta^l_{23}\leq 0$) 
for $\delta^l_{23}$ and relevant plots are depicted in Fig.1. From these plots it
is clear that negative values of $\cos\delta^l_{23}$, corresponding to 
$(\pi/2)\leq \delta^l_{23}\leq (3\pi/2)$, are preferable to accommodate best fist value
of $U^{MNS}_{\tau 3}$. In rest of the analyzes we will consider this range for
$\delta^l_{23}$. Again for negative $\cos\delta^l_{23}$, sine of $\delta^l_{23}$
can be positive ($(\pi/2)\leq \delta^l_{23}< \pi$) as well as negative
($\pi< \delta^l_{23}\leq (3\pi/2)$). From Eq.(36) it can be seen that, as both 
$\lambda_{23}$ and $\lambda_{13}$ are positive, the relative sign between 
$\sin\delta^l_{23}$ and $\sin\delta^l_{13}$ should be negative in order to have
$\sin\theta^{\nu}_{13}>0$.   \\
\begin{figure}[h!]
\centering
\subfloat[]{\includegraphics[scale=0.42]{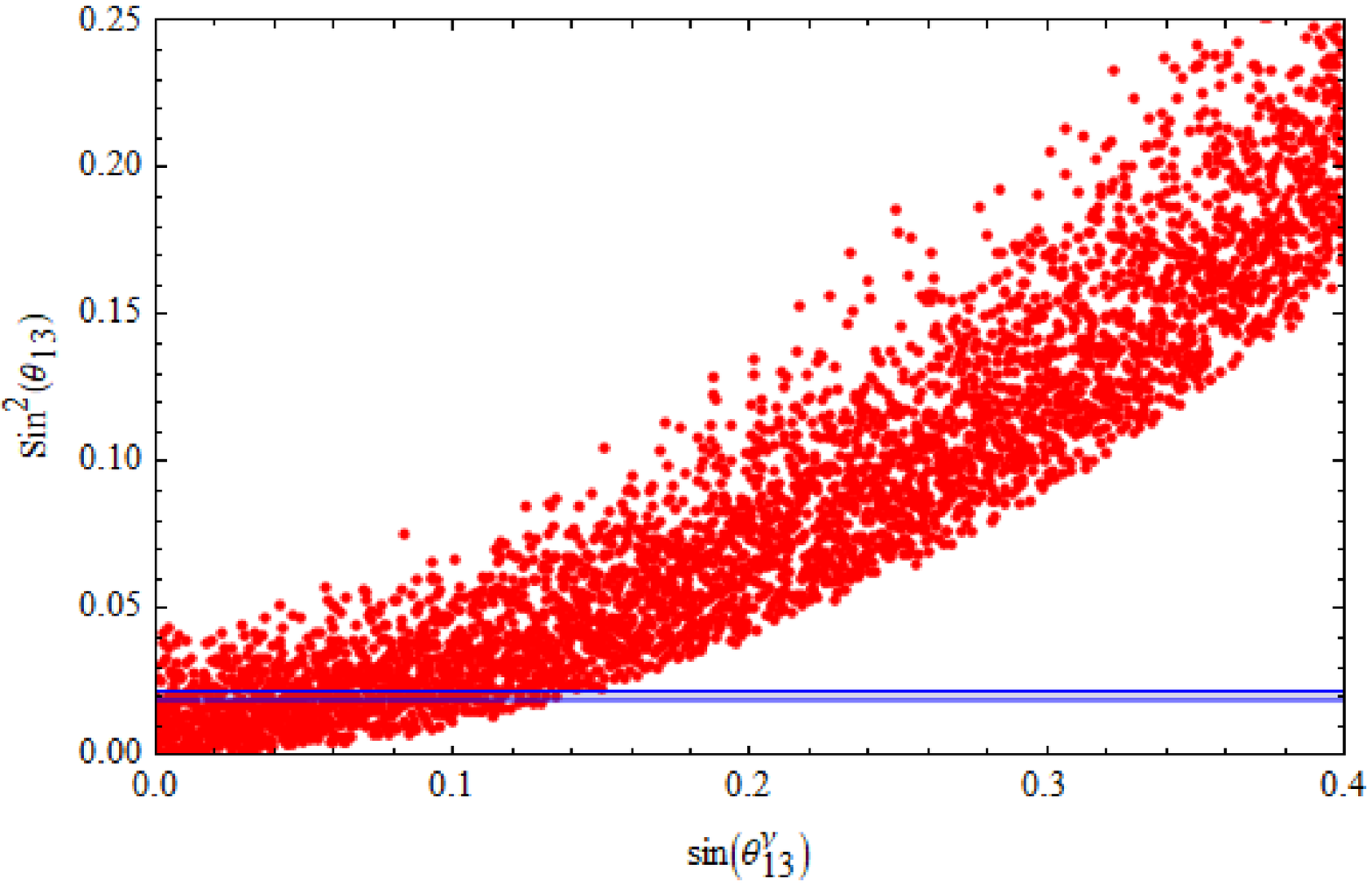}}
\subfloat[]{\includegraphics[scale=0.42]{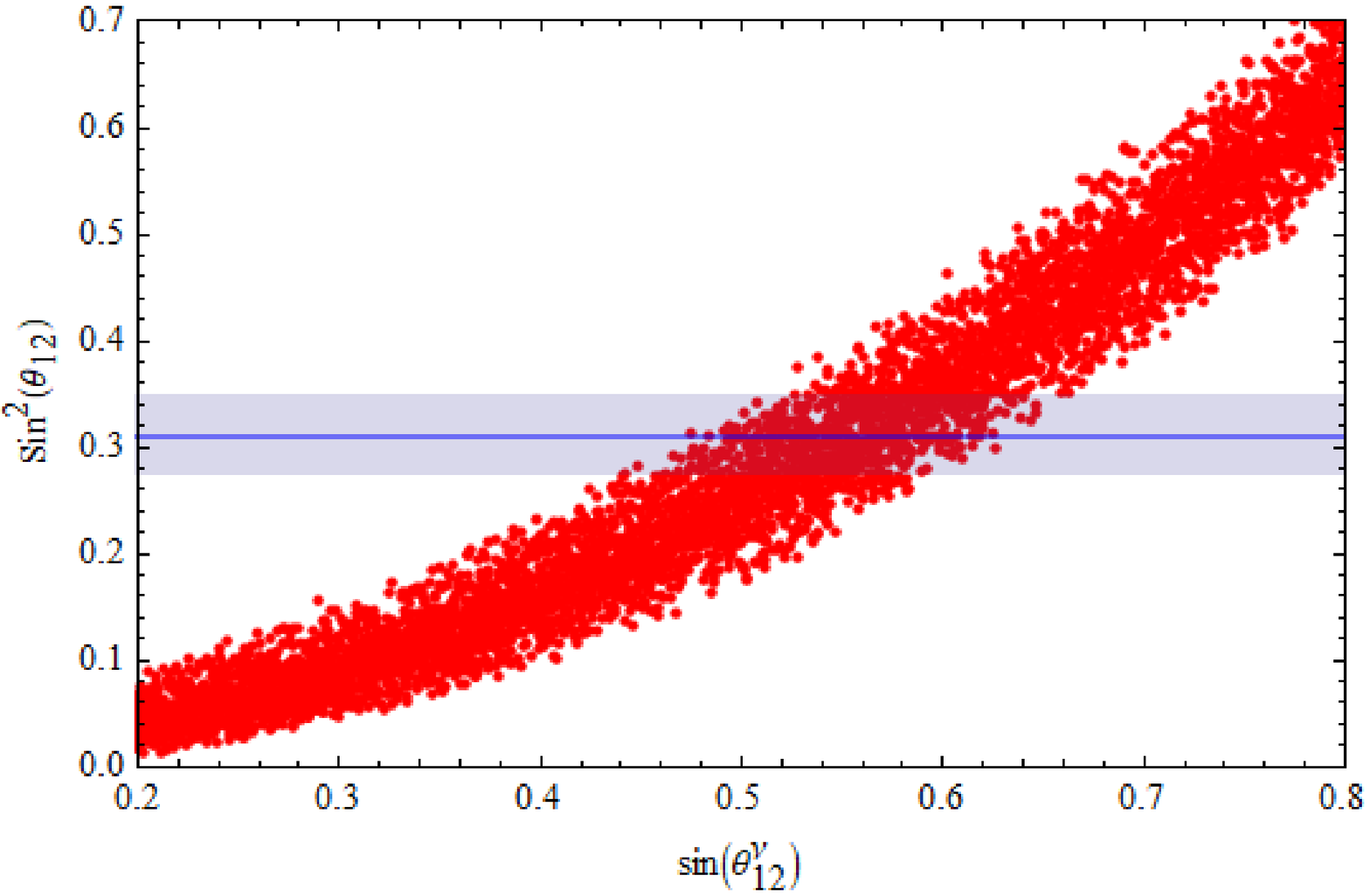}}
\caption{(a) Correlation between $\sin^2\theta_{13}$ and $\sin\theta^{\nu}_{13}$. 
Horizontal blue band represents the $3\sigma$ range of $\sin^2\theta_{13}$. 
(b) Correlation between $\sin^2\theta_{12}$ and $\sin\theta^{\nu}_{12}$. Horizontal 
colored band and the blue line represent the $3\sigma$ range and the best fit value
 of $\sin^2\theta_{12}$ respectively.}
\end{figure}
\begin{figure}[h!]
\centering
\subfloat[]{\includegraphics[scale=0.42]{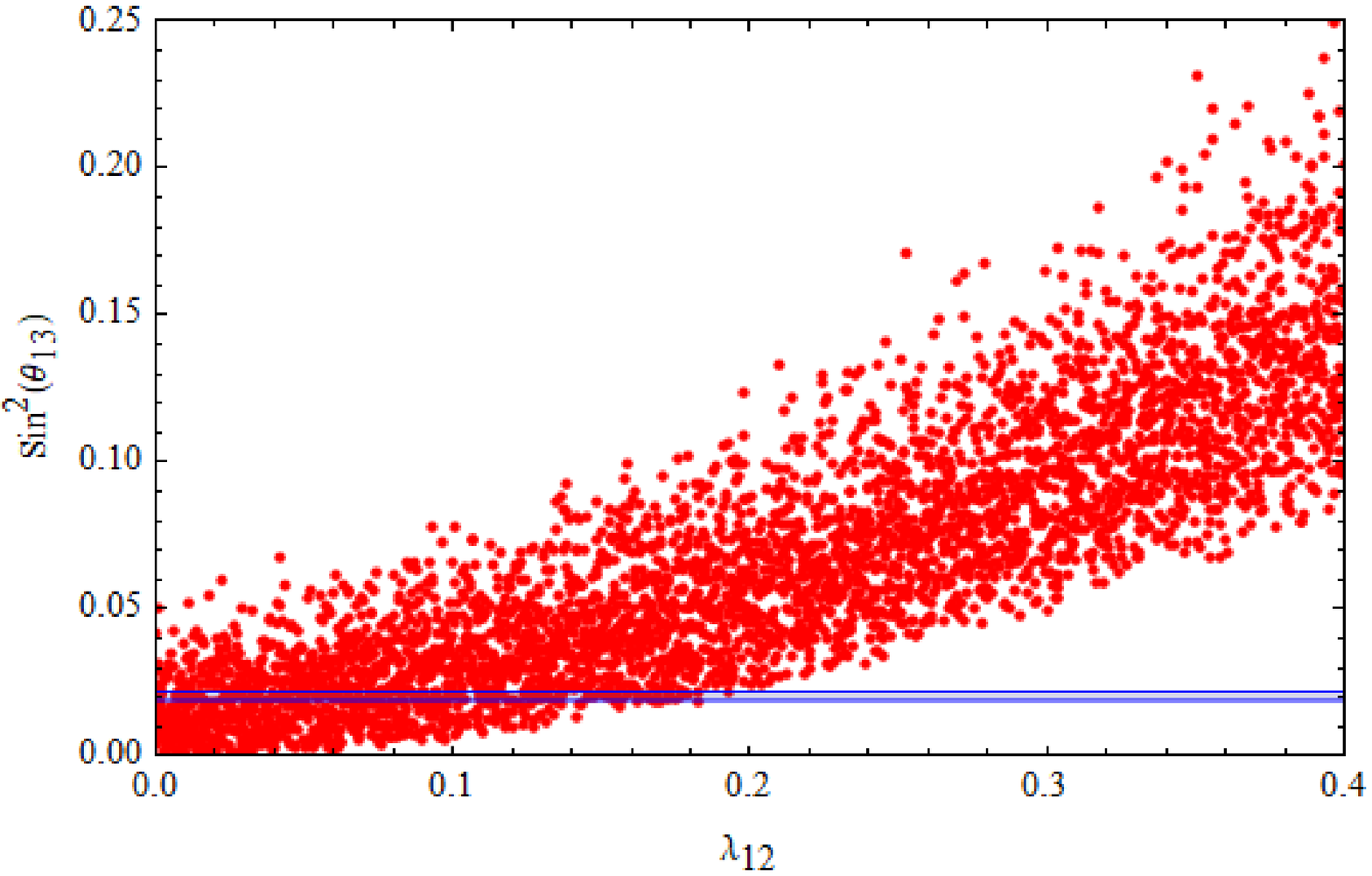}}
\subfloat[]{\includegraphics[scale=0.42]{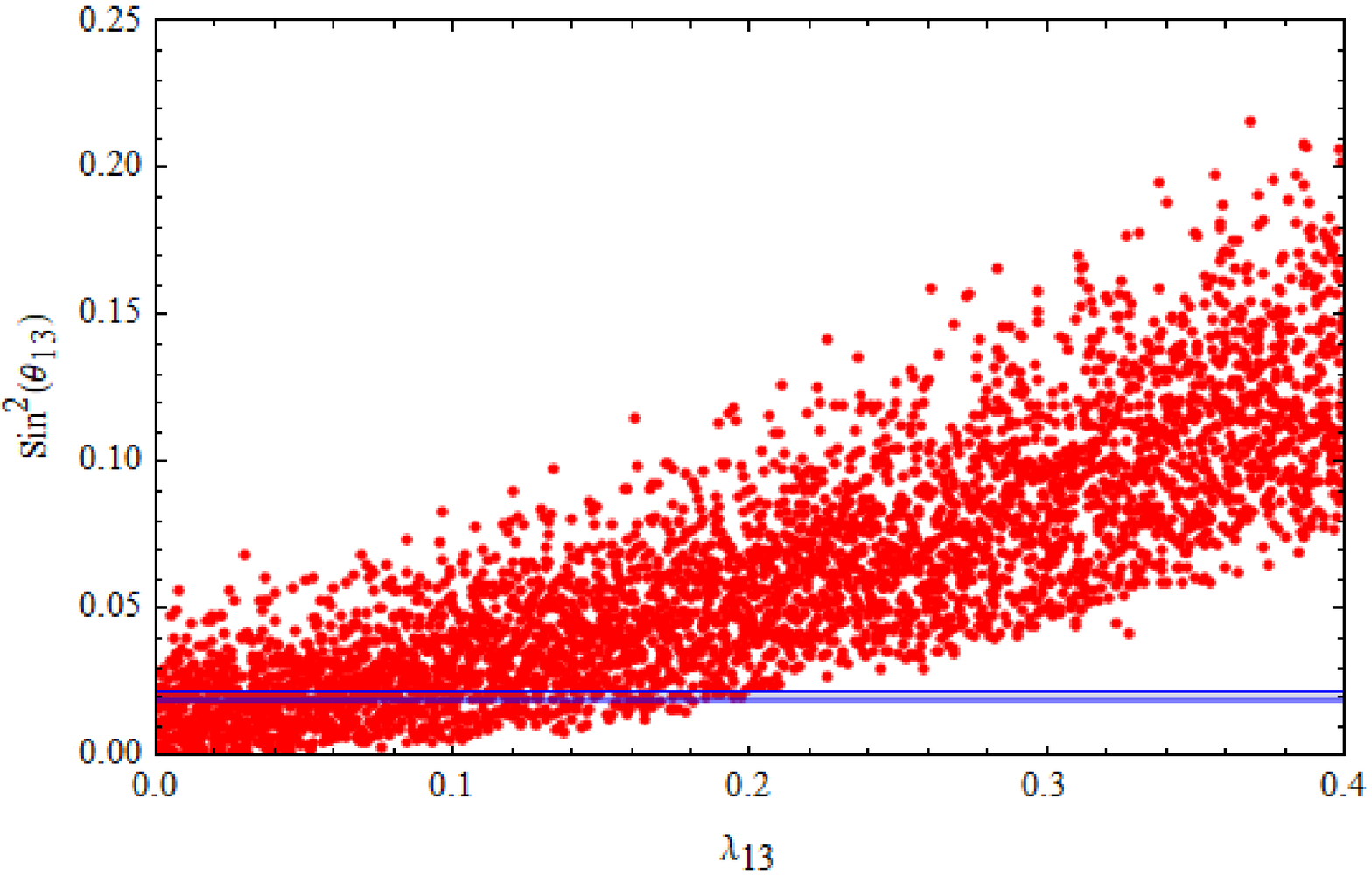}}
\caption{(a) Correlation between $\sin^2\theta_{13}$ and $\lambda_{12}$ and 
(b) that of $\sin^2\theta_{13}$ and $\lambda_{13}$. Horizontal blue bands represent 
the $3\sigma$ range of $\sin^2\theta_{13}$. }
\end{figure}                                   
\indent From Eq.(32) we get 
\begin{align}
\sin^2\theta_{13} & \simeq \sin^2\theta^{\nu}_{13} -\sqrt{2}\sin\theta^{\nu}_{13}
                            \left( \lambda_{12}\cos\delta^l_{12}+\lambda_{13}\cos\delta^l_{13}\right) 
                                                                           \nonumber \\
                          & \ \ \ \ + \frac{1}{2}\left(\lambda_{12}^2+\lambda_{13}^2 
                            +2\lambda_{12}\lambda_{13}\cos(\delta^l_{12}-\delta^l_{13}) \right).
\end{align} 
The sine of other two mixing angles can be obtained using Eq.(6) and are 
given by
\begin{align}
\sin^2\theta_{23} & \simeq \frac{1}{2} -\lambda_{23}\cos\delta^l_{23} 
                          +\frac{1}{\sqrt{2}}\sin\theta^{\nu}_{13}
                             \left( \lambda_{12}\cos\delta^l_{12} -\lambda_{13}\cos\delta^l_{13}\right)
                                                          \nonumber \\  
                   & \ \ \ \ \ \ -\frac{1}{4}\left(\lambda_{12}^2-\lambda_{13}^2 \right)  
                                -\frac{1}{2}\lambda_{12}\lambda_{13}\cos(\delta^l_{12}-\delta^l_{13}),                                   
\end{align}
\begin{align}
\sin^2\theta_{12} & \simeq \sin^2\theta^{\nu}_{12} \mp \frac{1}{\sqrt{2}} \sin 2\theta^{\nu}_{12}
                           \left( \lambda_{12}\sin\delta^l_{12} -\lambda_{13}\sin\delta^l_{13}\right)        
                                                                       \nonumber \\
                    & \ \ \ \ +\left(\frac{1}{2} -\sin^2\theta^{\nu}_{12} \right)                                           
                         \left[ \lambda_{12}^2+\lambda_{13}^2
                             -2\lambda_{12}\lambda_{13}\cos(\delta^l_{12}-\delta^l_{13}) \right]  
                                                                       \nonumber \\
                   & \ \ \ \  \mp \frac{1}{\sqrt{2}} \sin 2\theta^{\nu}_{12}
                            \left[ \lambda_{12}\lambda_{23}\sin(\delta^l_{12}+\delta^l_{23})
                                    -\lambda_{23}\lambda_{13}\sin(\delta^l_{23}-\delta^l_{13}) \right].                                    
\end{align} 
In the above expressions we have considered the terms up to second order in 
$\lambda_{ij}$. The '$\mp$' signs in the second term of right hand side of
Eq.(40) correspond to Case I and Case II respectively.                                   
The correlation between $\sin^2\theta_{13}$ and $\sin\theta^{\nu}_{13}$ is shown 
in Fig.  2(a) where the best fit value of $\sin\theta_{13}=0.149$ (Table 1). From
this plot we can see that the allowed values of $\sin\theta^{\nu}_{13}$ are less
than $\sin\theta_{13}$ as expected. The correlation between $\sin^2\theta_{12}$ 
and $\sin\theta^{\nu}_{12}$ (Fig. 2(b)) shows a positive linear relationship.
An allowed range of $0.5-0.6$ can be read off the correlation plot for 
$\sin\theta^{\nu}_{12}$ corresponding  to the best fit value of $\sin^2\theta_{12}$ ($0.556$). 
We have also studied the correlation between $\sin^2\theta_{13}$ and
$\lambda_{12}$ and $\lambda_{13}$ and are presented in Fig. 3(a) and 3(b) 
respectively. These plots reflects that the values of the free parameters 
$\lambda_{12}$ and $\lambda_{13}$ may be taken closed to $0.1$. Similar observation can
also be made for $\lambda_{23}$ from Fig. 1(a). In these correlation studies
negative values of both $\cos\delta^l_{12}$ and $\cos\delta^l_{13}$ are found
to be preferable along with negative $\cos\delta^l_{23}$.  \\
\indent From the real and imaginary parts of Eq.(32) we can express the Dirac CP phase
$\delta $ of the lepton mixing matrix in terms of the free parameters. It is given by 
\begin{equation}
\tan\delta =\frac{\lambda_{12}\sin\delta_{12}+\lambda_{13}\sin\delta_{13}
                         -\lambda_{12}\lambda_{23}\sin(\delta^l_{12}+\delta^l_{23})
                           -\lambda_{23}\lambda_{13}\sin(\delta^l_{23}-\delta^l_{13})}
                        {\sqrt{2}s^{\nu}_{13}-\lambda_{12}\cos\delta_{12}-\lambda_{13}\cos\delta_{13}
                         +\lambda_{12}\lambda_{23}\cos(\delta^l_{12}+\delta^l_{23})
                           -\lambda_{23}\lambda_{13}\cos(\delta^l_{23}-\delta^l_{13})}.
\end{equation}
The Jarlskog invariant can be obtained using Eq.(7) and is found to be  
\begin{align}
J & \simeq \pm \frac{1}{2}s^{\nu}_{12}c^{\nu}_{12}s^{\nu}_{13}\left(c^{\nu}_{13}\right)^2
              \mp \frac{1}{2\sqrt{2}} s^{\nu}_{12}c^{\nu}_{12}\left(c^{\nu}_{13}\right)^3   
                 \left( \lambda_{12}\cos\delta^l_{12} +\lambda_{13}\cos\delta^l_{13}\right)
                                                                \nonumber \\
  & \ \ \ \ +\frac{1}{2\sqrt{2}} \left[ \left(s^{\nu}_{12}\right)^2\left(c^{\nu}_{13}\right)^2
                      -\left(c^{\nu}_{12}\right)^2 \right] s^{\nu}_{13}c^{\nu}_{13}
                     \left( \lambda_{12}\sin\delta^l_{12} -\lambda_{13}\sin\delta^l_{13}\right)
                                                                \nonumber \\
  & \ \ \ \  \pm \frac{1}{2\sqrt{2}} s^{\nu}_{12}c^{\nu}_{12}\left(c^{\nu}_{13}\right)^3
                     \left[ \lambda_{12}\lambda_{23}\cos(\delta^l_{12}-\delta^l_{23})
                            - \lambda_{23}\lambda_{13}\cos(\delta^l_{23}+\delta^l_{13}) \right]
                                                                \nonumber \\
   & \ \ \ \  -\frac{1}{2}\left[ \left(s^{\nu}_{12}\right)^2\left(c^{\nu}_{13}\right)^2
                            -\left(c^{\nu}_{12}\right)^2 \right] \left(c^{\nu}_{13}\right)^2
                                  \lambda_{12}\lambda_{13}\sin(\delta^l_{12}-\delta^l_{13}),                                                                                                                                                                                                                                                                                                                                                             
\end{align}
where '$\pm$' signs in the first term of right hand side correspond to Case I and 
Case II respectively. \\
\indent Fig. 4(a) shows the correlation between $\tan\delta$ and $\sin\theta^{\nu}_{13}$.
As per the global best fit value $\tan\delta=0.7$ (Table 1), we see that the predictions of
$\tan\delta$ are consistent with the observed data within the allowed range of
$\sin\theta^{\nu}_{13}$ ($0-0.15$). The correlation between $J$ and 
$\sin\theta^{\nu}_{13}$ (Fig 4(b)) is also consistent with the global analysis 
data \cite{global1} which estimates a maximal possible value of $0.033$ for the 
Jarlskog invariant. \\
\begin{figure}[]
\centering
\subfloat[]{\includegraphics[scale=0.42]{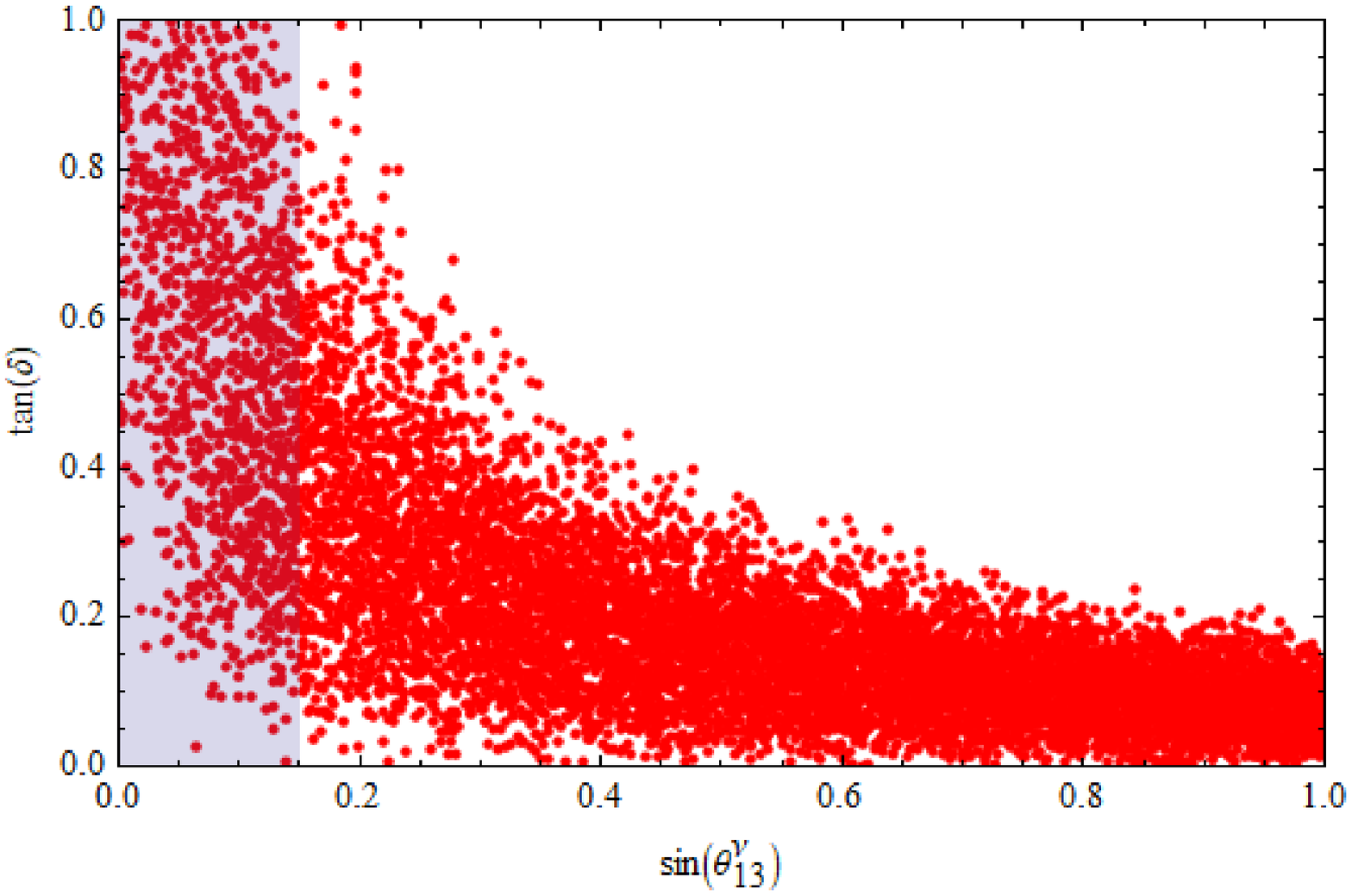}}
\subfloat[]{\includegraphics[scale=0.42]{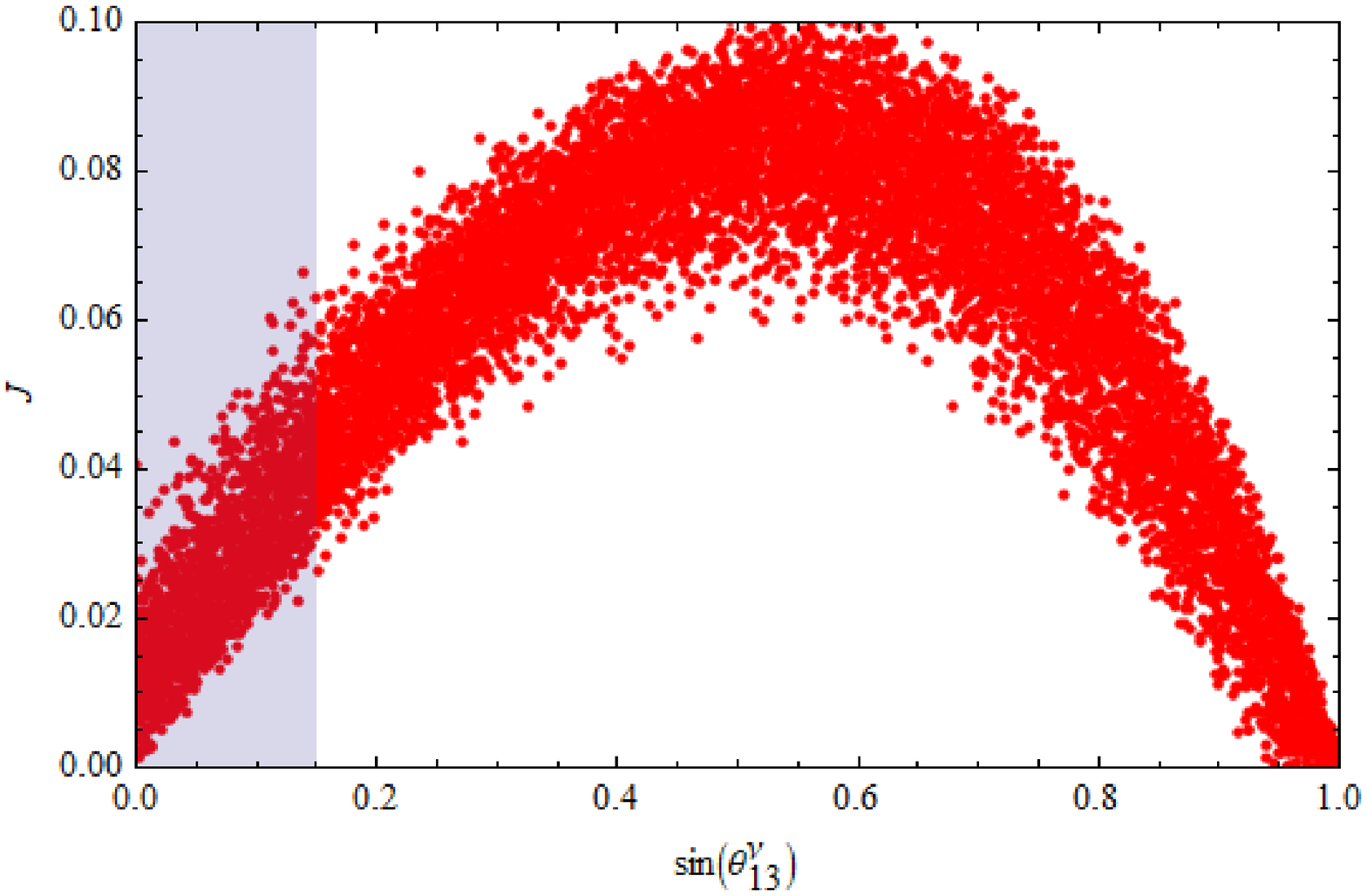}}
\caption{(a) Correlation between $\tan\delta$ and $\sin\theta^{\nu}_{13}$ 
and (b) that of $J$ and $\sin\theta^{\nu}_{13}$. Vertical blue bands represent the
allowed range of $\sin\theta^{\nu}_{13}$.}
\end{figure}
\indent A numerical analysis considering different possible input values for the 
eight free parameters is in order. To perform the analysis we first choose the values of the 
two mixing angles $\theta^{\nu}_{13}$ and $\theta^{\nu}_{12}$ of the neutrino 
mixing matrix $U^{\nu}$. We analyze two different cases with $\sin\theta^{\nu}_{13}=0.05$ 
and $0.1$. For both the cases neutrino solar angle is taken to be 
$\sin\theta^{\nu}_{12}=0.55$. To choose the values of the free parameters of $U^l$ 
we start with Eq.(36). As reflected from the correlation studies, we are interested 
in the values of $\lambda_{ij}$'s that are lying near to $0.1$. We run the analysis 
considering three types of values for $\lambda_{23}$ and $\lambda_{13}$ :
$\lambda_{23}<\lambda_{13}$, $\lambda_{23}>\lambda_{13}$ and
$\lambda_{23}=\lambda_{13}$. Specifically we have considered the values- 
$0.05/0.1$, $0.1/0.05$ and $0.1/0.1$ for $\lambda_{23}/\lambda_{13}$ for a given 
value of $\sin\theta^{\nu}_{13}$. After having the values of $\sin\theta^{\nu}_{13}$, 
$\lambda_{23}$ and $\lambda_{13}$ fixed,
Eq.(36) allows us to choose some possible values of $\sin\delta^l_{23}$ and 
$\sin\delta^l_{13}$. It is seen that with the chosen values of $\sin\theta^{\nu}_{13}$, 
$\lambda_{23}$ and $\lambda_{13}$, Eq.(36) constrains the values of $\sin\delta^l_{23}$ 
and $\sin\delta^l_{13}$ to a very narrow range. Remaining two free parameters  
($\lambda_{12}$ and $\delta^l_{12}$) can finally be chosen using Eq.(37). For a 
given set of values of $\sin\theta^{\nu}_{13}$ and $\lambda_{23}$, $\lambda_{13}$, 
$\delta^l_{23}$ and $\delta^l_{13}$, we vary the values of $\lambda_{12}$ in steps 
of $0.05$ and the corresponding values of $\sin\delta^l_{12}$
are solved from Eq.(37).     \\
\newcommand{\sheptexta}{$\sin\theta^{\nu}_{13}=0.05; \ \sin\theta^{\nu}_{12}=0.55$}
\newcommand{\sheptextb}{$\lambda_{23}<\lambda_{13}$}
\newcommand{\sheptextc}{$\lambda_{23}>\lambda_{13}$}
\newcommand{\sheptextd}{$\lambda_{23}=\lambda_{13}$}
\begin{table}[]
\centering
 \begin{tabular}{ccccccccccc}
\hline \hline
                     \multicolumn{11}{c}{\sheptexta}  \\
\hline \hline
    $\lambda_{12}$  &  $\lambda_{23}$ & $\lambda_{13}$ & $\delta^l_{12}/^{\circ}$ 
                         &  $\delta^l_{23}/^{\circ}$ & $\delta^l_{13}/^{\circ}$  
                               & $\sin^2\theta_{12}$ & $\sin^2\theta_{23}$ & $\sin^2\theta_{13}$ 
                                                      & $\tan\delta$ &  $J$   \\
\hline \hline
      \multicolumn{3}{c}{\sheptextb}  &  &  &  &  &  &  &  &  \\
\hline 
     $0.05$  &  $0.05$ &  $0.1$ & $245$ & $174$ & $225$ & $0.282$ & $0.550$ & $0.019$ & $0.717$ & $0.0252$   \\
     $0.1$  &  $0.05$ &  $0.1$ & $225$ & $174$ & $225$ & $0.298$ & $0.545$ & $0.032$ & $0.663$ & $0.0328$   \\
     $0.15$  &  $0.05$ &  $0.1$ & $218$ & $174$ & $225$ & $0.313$ & $0.538$ & $0.047$ & $0.633$ & $0.0395$   \\
\hline \hline
      \multicolumn{3}{c}{\sheptextc}  &  &  &  &  &  &  &  &  \\
\hline 
     $0.05$  &  $0.1$ &  $0.05$ & $196$ & $179$ & $196$ & $0.300$ & $0.598$ & $0.014$ & $0.170$ & $0.0260$  \\
     $0.1$  &  $0.1$ &  $0.05$ & $191$ & $179$ & $196$ & $0.304$ & $0.594$ & $0.025$  & $0.158$ & $0.0342$  \\ 
     $0.15$  &  $0.1$ &  $0.05$ & $190$ & $179$ & $196$ & $0.309$ & $0.589$ & $0.038$ & $0.150$ & $0.0421$  \\
\hline \hline
      \multicolumn{3}{c}{\sheptextd}  &  &  &  &  &  &  &  &  \\ 
\hline
     $0.05$  &  $0.1$ &  $0.1$ & $246$ & $177$ & $226$ & $0.278$ & $0.601$ & $0.018$ & $0.730$ & $0.0242$  \\
     $0.1$  &  $0.1$ &  $0.1$ & $225$ & $177$ & $226$ & $0.293$ & $0.595$ & $0.032$ & $0.671$ & $0.0321$  \\
     $0.15$  &  $0.1$ &  $0.1$ & $219$ & $177$ & $226$ & $0.306$ & $0.589$ & $0.048$ & $0.640$ & $0.0390$  \\
\hline  \hline   
\end{tabular}
\caption{Input values of the free parameters of $U^l$ and $U^{\nu}$ and 
corresponding predictions of the PMNS matrix parameters. }    
\end{table}
\newcommand{\sheptextf}{$\sin\theta^{\nu}_{13}=0.1; \ \sin\theta^{\nu}_{12}=0.55$}
\begin{table}[]
\centering
 \begin{tabular}{ccccccccccc}
\hline \hline
                     \multicolumn{11}{c}{\sheptextf}  \\
\hline \hline
    $\lambda_{12}$  &  $\lambda_{23}$ & $\lambda_{13}$ & $\delta^l_{12}/^{\circ}$ 
                         &  $\delta^l_{23}/^{\circ}$ & $\delta^l_{13}/^{\circ}$  
                          & $\sin^2\theta_{12}$  & $\sin^2\theta_{23}$ & $\sin^2\theta_{13}$ 
                                                      & $\tan\delta$ &  $J$   \\
\hline \hline
      \multicolumn{3}{c}{\sheptextc}  &  &  &  &  &  &  &  &  \\
\hline 
     $0.05$  &  $0.1$ &  $0.05$ & $227$ & $177$ & $227$ & $0.297$ & $0.598$ & $0.024$ & $0.349$ & $0.0322$   \\
     $0.1$  &  $0.1$ &  $0.05$ & $208$ & $177$ & $227$ & $0.305$ & $0.592$ & $0.039$ & $0.318$ & $0.0405$   \\
     $0.15$  &  $0.1$ &  $0.05$ & $203$ & $177$ & $227$ & $0.313$ & $0.586$ & $0.056$ & $0.300$ & $0.0479$   \\
\hline \hline
      \multicolumn{3}{c}{\sheptextd}  &  &  &  &  &  &  &  &  \\
\hline 
     $0.05$  &  $0.1$ &  $0.1$ & $244$ & $175$ & $218$ & $0.285$ & $0.603$ & $0.033$ & $0.440$ & $0.0359$  \\
     $0.1$  &  $0.1$ &  $0.1$ & $218$ & $175$ & $218$ & $0.294$ & $0.595$ & $0.051$  & $0.406$ & $0.0444$  \\ 
     $0.15$  &  $0.1$ &  $0.1$ & $210$ & $175$ & $218$ & $0.303$ & $0.587$ & $0.071$ & $0.386$ & $0.0514$  \\
\hline \hline   
\end{tabular}
\caption{Input values of the free parameters of $U^l$ and $U^{\nu}$ and 
corresponding predictions of the PMNS matrix parameters.}    
\end{table}
\begin{table}[]
\centering
 \begin{tabular}{ccccccccccc}
\hline \hline
                     \multicolumn{11}{c}{\sheptexta}  \\
\hline \hline
    $\lambda_{12}$  &  $\lambda_{23}$ & $\lambda_{13}$ & $\delta^l_{12}/^{\circ}$ 
                         &  $\delta^l_{23}/^{\circ}$ & $\delta^l_{13}/^{\circ}$  
                               & $\sin^2\theta_{12}$ & $\sin^2\theta_{23}$ & $\sin^2\theta_{13}$ 
                                                      & $\tan\delta$ &  $J$   \\
\hline \hline
  $0.05$  &  $0.05$ &  $0.1$ & $245$ & $174$ & $225$ & $0.324$ & $0.550$ & $0.019$ & $0.717$ & $-0.0250$   \\
  $0.05$  &  $0.1$ &  $0.05$ & $196$ & $179$ & $196$ & $0.304$ & $0.598$ & $0.014$ & $0.170$ & $-0.0261$  \\
  $0.05$  &  $0.1$ &  $0.1$ & $246$ & $177$ & $226$ & $0.328$ & $0.601$ & $0.018$ & $0.730$ & $-0.0241$  \\
\hline  \hline   
\end{tabular}
\caption{Input values of the free parameters of $U^l$ and $U^{\nu}$ and 
corresponding predictions of the PMNS matrix parameters. }    
\end{table}
\indent It is important to note that '$e3$', '$\mu 3$' and '$\tau 3$' elements of 
$U^{MNS}$ remain unaltered for the two cases : Case I and Case II, and thereby
predictions of $\sin^2\theta_{13}$ (Eq.(38)), $\sin^2\theta_{23}$ (Eq.(39)) and 
$\tan\delta$ (Eq.(41)) remain same for both the cases. However, other
elements of $U^{MNS}$ become different for the two cases due to the effect of
'$\pm$' signs
and consequently the expressions for $\sin^2\theta_{12}$ (Eq.(40)) and $J$
(Eq.(42)) suffer changes. Let us first consider Case I ($\theta^{\nu}_{23}=\pi/4$, 
$\delta^{\nu}= + \pi/2$). 
The different input values of the parameters of charged lepton mixing matrix $U^l$ 
with $\sin\theta^{\nu}_{13}=0.05$ and $\sin\theta^{\nu}_{12}=0.55$ and
corresponding predictions of the lepton mixing angles are presented in Table 2.
Similar numerical predictions with same type of input values of the parameters 
of $U^l$ are presented in Table 3 for $\sin\theta^{\nu}_{13}=0.1$ and 
$\sin\theta^{\nu}_{12}=0.55$.
Few significant remarks can be drawn from this numerical analysis regarding the 
connection between $\lambda_{ij}$'s and the lepton mixing angles. The value of 
the atmospheric mixing angle $\theta_{23}$ mainly depends on $\lambda_{23}$ while
the effects of the other two parameters ($\lambda_{12}$, $\lambda_{13}$) are
relatively small. It can also be observed from Eq.(39) where the first factor $1/2$
stands for the maximal value of $\sin^2\theta_{23}$. The value of $\lambda_{23}$
together with $\cos\delta^l_{23}$ accounts for the deviation of $\theta_{23}$
from the maximal value without any significant effect from $\lambda_{12}$ and 
$\lambda_{13}$. Further, since $\lambda_{23}$ is positive, the sign of $\cos\delta^l_{23}$
basically determines the octant for $\theta_{23}$. As per the indication  
$\sin^2\theta_{23}>0.5$, revealed by oscillation data, we constrain 
$\cos\delta^l_{23}$ to negative values in the present analysis. 
Eq.(38) shows that the prediction on reactor angle $\theta_{13}$ depends both on 
$\lambda_{12}$ and $\lambda_{13}$ and corresponding phases in the leading order 
while the effect of $\lambda_{23}$ is negligible. The same is also visible in
the numerical results.
The first block of Table 2 displays the predictions of the lepton mixing angles for 
the fixed value of $\lambda_{23}=0.05$ while those of second and third blocks
correspond to $\lambda_{23}=0.1$. The predictions of $\sin^2\theta_{23}$ 
corresponding to $\lambda_{23}=0.05$ are less than the global best fit value
($0.58$) while those corresponding to $\lambda_{23}=0.1$ are obtained at the
desired level. We can see that while varying $\lambda_{12}$ for a fixed
($\lambda_{23}/\lambda_{13}$), $\sin^2\theta_{23}$ does not change significantly.
This suggests that value of $\lambda_{23}$ closed to $0.1$ is suitable to
generate the observed atmospheric mixing angle in the second octant. 
The prediction of $\sin^2\theta_{13}$ depends both on $\lambda_{12}$ or $\lambda_{13}$.
From Tables 2 and 3 we can see that for a fixed value of $\lambda_{23}$,
$\sin^2\theta_{13}$ gradually increases either with $\lambda_{12}$ or $\lambda_{13}$.
As per the global best fit value of $\sin^2\theta_{13}=0.022$, we observe that
the values of $\lambda_{12}\sim 0.05$ (or $0.1$) and $\lambda_{13}\sim 0.1$ 
(or $0.05$) may serve the desired purpose. However, as the orders of 
$\sin\theta^{\nu}_{13}$ and $\lambda_{ij}$'s are same, relative magnitudes of 
$\sin\theta^{\nu}_{13}$, $\lambda_{12}$ and $\lambda_{13}$  
play certain role in the prediction of $\sin^2\theta_{13}$. A comparative analysis
on the prediction of $\sin^2\theta_{13}$ can be made from Table 2 and Table 3. We
find that the predictions of $\sin^2\theta_{13}$ are relatively higher for 
$\sin\theta^{\nu}_{13}=0.1$ compared to $\sin\theta^{\nu}_{13}=0.05$. Hence the 
choices $\sin\theta^{\nu}_{13}<0.1$ are preferable in predicting the third
lepton mixing angle $\theta_{13}$ at its global best fit value for input values
of $\lambda_{12}, \lambda_{13}\sim 0.1$.    
 The prediction on solar angle $\theta_{12}$ basically
depends on $\theta^{\nu}_{12}$ with small perturbation from the small 
$\lambda_{ij}$'s. We first analyze the predictions taking input values for 
$\sin\theta^{\nu}_{12}<0.55$ where $0.55$ is the global
best fit value of $\sin\theta_{12}$. As the predictions of $\sin^2\theta_{12}$
are found to be less than that of the global best fit ($0.31$), we compute the analysis
assuming $\sin\theta^{\nu}_{12} \approx \sin\theta_{12}=0.55$. Corresponding 
results can be read from Table 2 and Table 3. \\
\indent Turning to the prediction of the Dirac CP violation effects, we note
that the global analysis of $3\nu$ oscillation data \cite{global1} provides a best 
fit value of $\tan\delta=0.7$ for normal hierarchy (Table 1). Further, it 
indicates a value of the Jarlskog invariant $J=-0.019$ for non-maximal mixing.
In view of these predictions we can compare the results of Table 2 and Table 3
and we find that the choice $\sin\theta^{\nu}_{13}=0.05$ is more suitable.
It is also reflected from the correlation plot of $J$ and $\sin\theta^{\nu}_{13}$ 
(Fig. 4(b)) that lower values of $\sin\theta^{\nu}_{13}$ correspond to smaller 
values of $J$. Regarding the prediction of $\tan\delta$, we however notice
a difference in the second block of Table 2 where the predicted
values are significantly small as compared to the global best fit.  \\
\indent In view of the overall prediction of the lepton mixing parameters we 
can find particular interest in the results of Table 2 obtained for 
$\sin\theta^{\nu}_{13}=0.05$ and $\sin\theta^{\nu}_{12}=0.55$. We can see that
the input values $(\lambda_{12},\lambda_{23},\lambda_{13})=(0.1,0.1,0.05)$
(second block) can generate the three mixing angles at the desired best fit 
values. However the prediction on $\tan\delta$ is not consistent with the
global best fit value. Moving to the third block, the input values 
$(\lambda_{12},\lambda_{23},\lambda_{13})=(0.05,0.1,0.1)$ are also good in
predicting all the parameters at desired level except for the solar angle which
lies below the global best fit value. Although the prediction of $\sin^2\theta_{12}$ 
can be lifted by increasing the value of $\lambda_{12}$, it, in turn, affects 
the prediction of $\sin^2\theta_{13}$. It is interesting to note that
the best fit value of $\sin^2\theta_{12}$ can be maintained without
affecting the prediction of $\sin^2\theta_{13}$ in the other case (Case II). 
Numerical predictions with different input values for Case II are presented
in Table 4. We can compare the predictions of the mixing parameters
in the two cases for the input values 
$(\lambda_{12},\lambda_{23},\lambda_{13})=(0.05,0.1,0.1)$ from Table 2 and
Table 4 which shows that prediction on $\sin^2\theta_{12}$ rises from $0.278$
to $0.328$.

\section{Summary and conclusion}

\indent The role of $\mu-\tau$ reflection symmetry, as it features a non zero $\theta_{13}$ 
besides maximal $\theta_{23}$, is significant in the study of lepton flavour models.
In this work we point out that the reflection symmetric nature of the lepton mixing
matrix is not reflected back while substituting the maximal values of $\theta_{23}$
and $\delta$ in the standard parametrization. Motivated by this observation we look
for possible solution to this ambiguity and find that the symmetry can be restored
by assigning maximal values of the Majorana phases as well in addition to maximal Dirac phase
$\delta$. A noteworthy contribution from the un-physical phases is remarked in the
symmetry realization. \\
\indent We have exercised the scenario under a broken symmetry such
that deviated values of maximal $\theta_{23}$ and maximal $\delta$ can be accommodated.
For this purpose, contributions from charged lepton sector is considered as a possible 
scheme to generate the broken symmetry. We implant the reflection symmetry in the neutrino
mixing matrix $U^{\nu}$ and study the consequences in the lepton mixing matrix under the
basis where the charged lepton mass matrix is non diagonal. The perturbations from
the charged lepton sector are assumed to be small and the mixing matrix $U^l$ is
parametrized in terms of the small parameters $\lambda_{ij}$'s ($ij=12,\ 23,\ 13$). 
The parametrization of $U^l$, in addition, includes three complex phases 
$\delta^l_{12}$, $\delta^l_{23}$ and $\delta^l_{13}$. The predictions of the lepton 
mixing parameters of $U^{MNS}$ depend on the six free parameters of $U^l$ along with 
$\theta^{\nu}_{13}$ and $\theta^{\nu}_{12}$ of $U^{\nu}$. A comprehensive numerical 
analysis is made in this work regarding the possible values of the free parameters 
in determining desired lepton mixing
parameters. We primarily focus on the possible values of the parameters 
$\lambda_{ij}$'s and the two neutrino mixing angles. The values of $\lambda_{ij}$'s
are found to be in the order of $0.1$ to generate the lepton mixing parameters
consistent with global data. It is apparent that
the lepton mixing angles are close to the values of the corresponding neutrino mixing
angles with small perturbations from $\lambda_{ij}$'s. Reflection symmetry fixes the 
neutrino atmospheric angle at the maximal value 
$\sin^2\theta^{\nu}_{23}=1/2$. As per
the global best fit value of $\sin^2\theta_{23}=0.58$, we expect a slight 
positive contribution from the charged lepton mixing parameters. In the charged 
lepton correction scheme considered, $\lambda_{23}$ along with the associating 
phase $\delta^l_{23}$ is found to play the chief role in this regard. Values of
$\lambda_{23}\lesssim 0.1$ and negative values of $\cos\delta^l_{23}$ are suitable
to generate the global best fit value of $\sin^2\theta_{23}$. For the neutrino solar
angle, values of $\sin\theta^{\nu}_{12} \lesssim \sin\theta_{12}$ are found to
provide the desired predictions. There exists a distinction between the predictions
of $\sin^2\theta_{12}$ in the two separate cases (Case I and Case II). Prediction 
of $\sin^2\theta_{12}$ is better with respect to the global best fit value in
Case II ($\theta^{\nu}_{23}=\pi/4$, $\delta^{\nu}= -\pi/2$).
Prediction of $\sin^2\theta_{13}$ depends on 
both $\sin\theta^{\nu}_{13}$ and the parameters $\lambda_{12}$ and $\lambda_{13}$     
under the same order. It is observed that values of $\sin\theta^{\nu}_{13}\ll 0.15$
(global best fit value of $\sin\theta_{13}$) are more comfortable corresponding 
to the input values of $\lambda_{12},\lambda_{13}\sim 0.1$. Such small values of  
$\sin\theta^{\nu}_{13}$ are also preferable in predicting small non maximal values 
of the Jarlskog invariant as indicated by global analysis data. 
The Dirac CP violation effect is observed in terms of $\tan\delta$. It's prediction
is also found to be in nice agreement with the global best fit value and comfortable
 with the predictions of the three mixing angles. 
In conclusion, the charged lepton correction scheme studied can accommodate the 
current global analysis data with nice precision.   

\section*{Acknowledgment}
\indent The author would like to thank Dr. Subhankar Roy of Gauhati University, Guwahati, for
his help in fixing some technical difficulties related to this work.  

\appendix
\section{Equivalent parametrizations}
\numberwithin{equation}{section}
\setcounter{equation}{0}
\indent Here we have shown the equivalence between two parametrizations 
of mixing matrices that are related to this work. Let us consider the parametrization 
defined in Eq.(4). Hermitian conjugation of the lepton mixing matrix is 
\begin{equation}
U^{\dagger} = P^{\dagger}_2V^{\dagger} P^{\dagger}_1,
\end{equation} 
where 
\begin{equation}
P^{\dagger}_1 = \begin{pmatrix}
       e^{-i\phi_1} & 0 & 0 \\
        0 &  e^{-i\phi_2} & 0  \\
        0 & 0 &  e^{-i\phi_3}  \\
       \end{pmatrix}, \ \ \  P^{\dagger}_2 = \begin{pmatrix}
                                   e^{-i\alpha} & 0 & 0 \\
                                    0 &  e^{-i\beta} & 0  \\
                                    0 & 0 &  1  \\
                                     \end{pmatrix},
\end{equation}
and the mixing matrix $V$ is defined in Eq.(5).
Expanding $V$ in terms of three rotation matrices we have
\begin{equation}
U^{\dagger} = P^{\dagger}_2 R^{\dagger}_{12}\tilde{U}^{\dagger}_{13}R^{\dagger}_{23} P^{\dagger}_1,
\end{equation}
with
\begin{equation}
R_{12}= \begin{pmatrix}
        c_{12}  & s_{12} & 0 \\
       -s_{12}  & c_{12} & 0 \\
        0  & 0 & 1 \\
        \end{pmatrix}, \ \ \ R_{23}= \begin{pmatrix}
                                     1  & 0 & 0 \\
                                     0  & c_{23} & s_{23} \\
                                       0  & -s_{23} & c_{23} \\
                                      \end{pmatrix}, 
\end{equation}
\begin{equation}
\tilde{U}_{13}= \begin{pmatrix}
        c_{13}  & 0 & s_{13}e^{-i\delta} \\
           0  & 1 & 0 \\
        -s_{13}e^{i\delta} & 0 & c_{13} \\
        \end{pmatrix}.
\end{equation}
Now, commuting the phase matrix $P^{\dagger}_1$ step by step to the right in Eq.(A.1), 
it can be shown that the parametrization in Eq.(A.1) is equivalent to 
\begin{equation}
U^{\dagger} = P^{\dagger}U^{\dagger}_{12}U^{\dagger}_{13}U^{\dagger}_{23},
\end{equation} 
where $P^{\dagger}=P^{\dagger}_2P^{\dagger}_1 $ and
\begin{equation}
U^{\dagger}_{12}= \begin{pmatrix}
        c_{12}  &  -s_{12}e^{i\delta_{12}} \\
        s_{12}e^{-i\delta_{12}}  & c_{12} & 0 \\
            0 & 0 & 1 \\
        \end{pmatrix}, \ \ \ U^{\dagger}_{13}= \begin{pmatrix}
                                                c_{13}  &  0 &  -s_{13}e^{i\delta_{13}} \\
                                                0       &  1  &  0  \\
                                                s_{13}e^{-i\delta_{13}} & 0  & c_{13}  \\
                                                 \end{pmatrix},
\end{equation}
\begin{equation}
U^{\dagger}_{23}= \begin{pmatrix}
                   1      &  0    &   0  \\
                   0    &  c_{23}  &  -s_{23}e^{i\delta_{23}} \\
                   0   &  s_{23}e^{-i\delta_{23}} & c_{23}  \\
                                                 \end{pmatrix}.
\end{equation}
The new phases defined in the matrices $U_{12}$,  $U_{13}$ and  $U_{23}$ are related to the
phases of the former parametrization as
\begin{equation}
\delta_{12} = \phi_1 - \phi_2, \ \ \delta_{13} =\delta- \phi_1+ \phi_3, \ \ \delta_{23} = \phi_2 - \phi_3.
\end{equation}  

\section{Higher order expressions for mixing parameters}
The expressions for the lepton mixing angles and the Jarlskog invariant
given in Eqs.(38), (39), (40) and (42) are approximated up to second order in
$\lambda_{ij}$ where higher order terms are neglected in view of the observation
 $O(\lambda_{ij})\sim O(0.1)$. Here we have added higher order terms in these 
expressions with $O(\lambda_{ij})$ increased by one. Before writing down the
expressions we define the following compact notations for convenience : 
\begin{gather}
\Lambda(12,13)^{(1)}_{c\pm/s\pm} = \lambda_{12}\cos\delta^l_{12} \pm \lambda_{13}\cos\delta^l_{13}/
                           \lambda_{12}\sin\delta^l_{12} \pm \lambda_{13}\sin\delta^l_{13},   \nonumber \\        
\Lambda(12,23)^{(2)}_{c\pm/s\pm} = \lambda_{12}\lambda_{23}\cos(\delta^l_{12} \pm \delta^l_{23})/
                       \lambda_{12}\lambda_{23}\sin(\delta^l_{12} \pm \delta^l_{23}),   \nonumber \\  
\Lambda(23,13)^{(2)}_{c\pm/s\pm} = \lambda_{23}\lambda_{13}\cos(\delta^l_{23} \pm \delta^l_{13})/
                       \lambda_{23}\lambda_{13}\sin(\delta^l_{23} \pm \delta^l_{13}),   \nonumber \\  
\Lambda(12,13)^{(2)}_{c\pm/s\pm} = \lambda_{12}\lambda_{13}\cos(\delta^l_{12} \pm \delta^l_{13})/
                         \lambda_{12}\lambda_{13}\sin(\delta^l_{12} \pm \delta^l_{13}),  \nonumber \\  
\Lambda^{(3)} = \left(\lambda_{12}^2-\lambda_{13}^2 \right) \lambda_{23}\cos\delta^l_{23}. 
\end{gather}
With these relevant expressions become
\begin{align}
\sin^2\theta_{13} & = \left(s^{\nu}_{13}\right)^2 -\sqrt{2}s^{\nu}_{13}
                       \left[ \Lambda(12,13)^{(1)}_{c+}-\Lambda(12,23)^{(2)}_{c+} 
                                         + \Lambda(23,13)^{(2)}_{c-} \right] \nonumber \\
                     & \ \ \ \ + \frac{1}{2}\left[\lambda_{12}^2+\lambda_{13}^2
                                 +2 \Lambda(12,13)^{(2)}_{c-}  \right] -\Lambda^{(3)} 
                                 +\frac{1}{2}\left( \lambda_{12}^2+\lambda_{13}^2 \right)\lambda_{23}^2 
                                                                 \nonumber \\
                     & \ \ \ \  -\lambda_{12}\lambda_{23}^2\lambda_{13}
                                    \cos(\delta^l_{12}+ 2\delta^l_{23}-\delta^l_{13}),
 \end{align}
\begin{align}
\sin^2\theta_{23} & \simeq \frac{1}{2} -\lambda_{23}\cos\delta^l_{23} 
                           -\frac{1}{4}\left[\lambda_{12}^2-\lambda_{13}^2 
                                          +2\Lambda(12,13)^{(2)}_{c-} \right]  \nonumber \\ 
                  & \ \ \ \ \  +\frac{1}{\sqrt{2}}s^{\nu}_{13}\left[ \Lambda(12,13)^{(1)}_{c-}
                               -\Lambda(12,23)^{(2)}_{c+}-\Lambda(23,13)^{(2)}_{c-} \right] \nonumber \\ 
                  & \ \ \ \ \  +\frac{1}{2} \left[ 2\sqrt{2}s^{\nu}_{13} \Lambda(12,13)^{(1)}_{c+}
                                    -\left(s^{\nu}_{13}\right)^2 -\lambda_{12}^2+\lambda_{23}^2\right]
                                     \lambda_{23}\cos\delta^l_{23} \nonumber \\                 
                   & \ \ \ \ \ -\Lambda(12,13)^{(2)}_{s-}\lambda_{23}\sin\delta^l_{23},                                                    
\end{align}
\begin{align}
\sin^2\theta_{12} & \simeq \sin^2\theta^{\nu}_{12} \left[ 1- \lambda_{12}^2-\lambda_{13}^2
                        +2\Lambda(12,13)^{(2)}_{c-} -2\Lambda^{(3)}\right] 
                        + \frac{1}{2}\left( \lambda_{12}^2+\lambda_{13}^2 \right) \nonumber \\
                  & \ \ \ \   - \Lambda(12,13)^{(2)}_{c-}+ \frac{1}{2}\Lambda^{(3)}
                           + \sqrt{2}\sin^2\theta^{\nu}_{12} s^{\nu}_{13}
                           \left[ \Lambda(12,23)^{(2)}_{c+} -\Lambda(23,13)^{(2)}_{c-} \right]  \nonumber \\
                  & \ \ \ \ \mp \frac{1}{\sqrt{2}} \sin 2\theta^{\nu}_{12}
                           \left[ \Lambda(12,13)^{(1)}_{s-} \left( 1+\Lambda(12,13)^{(2)}_{c-} \right)
                                 + \Lambda(12,23)^{(2)}_{s+} -\Lambda(23,13)^{(2)}_{s-} \right]
                                                                                     \nonumber \\
                   & \ \ \ \ \pm \frac{1}{2\sqrt{2}} \sin 2\theta^{\nu}_{12} 
                            \left[ \sqrt{2}s^{\nu}_{13} \left( \lambda^2_{12}\sin 2\delta^l_{12}
                                        -\lambda^2_{13}\sin 2\delta^l_{13} \right)
                               -\left(s^{\nu}_{13}\right)^2 \Lambda(12,13)^{(1)}_{s-} \right],                                                                                                                                                                               
\end{align}
\begin{align}
J & \simeq \pm \frac{1}{2}s^{\nu}_{12}c^{\nu}_{12}s^{\nu}_{13}\left(c^{\nu}_{13}\right)^2
                        \left[1-2\lambda^2_{12}-2\lambda^2_{23}-2\lambda^2_{13} 
                        -\Lambda(12,13)^{(2)}_{c-}-2\Lambda(12,13)^{(2)}_{c+} \right]  \nonumber \\
      & \ \ \ \  \pm \frac{1}{2\sqrt{2}}s^{\nu}_{12}c^{\nu}_{12}\left(s^{\nu}_{13}\right)^2 c^{\nu}_{13} 
                     \left( 3\lambda_{12}\cos\delta^l_{12} + \lambda_{13}\cos\delta^l_{13}\right) 
                                                                                    \nonumber \\
      & \ \ \ \  \mp \frac{1}{2\sqrt{2}}s^{\nu}_{12}c^{\nu}_{12}\left(c^{\nu}_{13}\right)^3
                    \left[ 1-\frac{3}{2}\lambda^2_{12}-\lambda^2_{23}-\frac{1}{2}\lambda^2_{13}\right]
                             \Lambda(12,13)^{(1)}_{c+}           \nonumber \\
      & \ \ \ \  \pm \frac{1}{2\sqrt{2}}s^{\nu}_{12}c^{\nu}_{12}\left(c^{\nu}_{13}\right)^3
                    \left[ \Lambda(12,23)^{(2)}_{c-} -\Lambda(23,13)^{(2)}_{c+}
                    -\left(\lambda^2_{12} -2\lambda^2_{13}\right)\lambda_{12}\cos\delta^l_{12}
                                                                   \right]  \nonumber \\  
      & \ \ \ \  +\frac{1}{2\sqrt{2}}\left(s^{\nu}_{12}\right)^2s^{\nu}_{13} \left(c^{\nu}_{13}\right)^3
                \left[ \Lambda(12,13)^{(1)}_{s-} +\Lambda(12,23)^{(2)}_{s-}
                                            -\Lambda(23,13)^{(2)}_{s+} \right]          \nonumber \\  
       & \ \ \ \  -\frac{1}{2\sqrt{2}}\left(c^{\nu}_{12}\right)^2s^{\nu}_{13}c^{\nu}_{13}
               \left[\Lambda(12,13)^{(1)}_{s-}-2\lambda_{12}\lambda_{23}\cos\delta^l_{12}\cos\delta^l_{23}
                            +\Lambda(23,13)^{(2)}_{s+} \right]          \nonumber \\
       & \ \ \ \  -\frac{1}{2}\left[\left(s^{\nu}_{12}\right)^2\left(c^{\nu}_{13}\right)^2
                              -\left(c^{\nu}_{12}\right)^2 \right] \left(c^{\nu}_{13}\right)^2
                                              \Lambda(12,13)^{(2)}_{s-}.                                                                                                                                                                                                                                                                                                                                                                                                                                                                                                                                                                                                                                                                                                                                                                                                                                
\end{align}


\end{document}